\documentclass[12pt]{article}

\usepackage{amssymb}
\usepackage{amsmath}
\usepackage{amscd}
\usepackage{latexsym}
\usepackage{graphicx}
\usepackage{url}

\usepackage{cite}

\newcommand{\be}{\begin{equation}}
\newcommand{\ee}{\end{equation}}
\newcommand{\Dlt}{\Delta}
\newcommand{\dlt}{\delta}
\newcommand{\prt}{\partial}

\newcommand{\bt}{\beta}

\newcommand{\al}{\alpha}
\newcommand{\ra}{\rightarrow}

\newcommand{\gm}{\gamma}

\newcommand{\lbd}{\lambda}
\newcommand{\Lbd}{\Lambda}

\begin{document}

\begin{center}

{\Large{\bf Dynamical system theory \\of periodically collapsing bubbles} \\ [5mm]

V.I. Yukalov$^{1,2,*}$, E.P. Yukalova$^{1,3}$, and
D. Sornette$^{1,4}$} \\ [3mm]

{\it
$^1$Department of Management, Technology and Economics, \\
ETH Z\"urich (Swiss Federal Institute of Technology) \\
Scheuchzerstrasse 7,  Z\"urich CH-8032, Switzerland \\ [3mm]

$^2$Bogolubov Laboratory of Theoretical Physics, \\
Joint Institute for Nuclear Research, Dubna 141980, Russia \\ [3mm]

$^3$Laboratory of Information Technologies, \\
Joint Institute for Nuclear Research, Dubna 141980, Russia \\ [3mm]

$^4$Swiss Finance Institute, c/o University of Geneva, \\
40 blvd. Du Pont d'Arve, CH 1211 Geneva 4, Switzerland}

\end{center}

\vskip 1.5cm

\begin{abstract}

We propose a reduced form set of two coupled continuous time equations
linking the price of a representative asset and the price of a bond, the
later quantifying the cost of borrowing. The feedbacks between asset prices 
and bonds are mediated by the dependence of their ``fundamental values'' on 
past asset prices and bond themselves. The obtained nonlinear self-referencing
price dynamics can induce, in a completely objective deterministic way, the 
appearance of periodically exploding bubbles ending in crashes. Technically, 
the periodically explosive bubbles arise due to the proximity of two types 
of bifurcations as a function of the two key control parameters $b$ and $g$, 
which represent, respectively, the sensitivity of the fundamental asset price 
on past asset and bond prices and of the fundamental bond price on past asset 
prices. One is a Hopf bifurcation, when a stable focus transforms into an
unstable focus and a limit cycle appears. The other is a rather unusual
bifurcation, when a stable node and a saddle merge together and disappear, 
while an unstable focus survives and a limit cycle develops. The lines, 
where the periodic bubbles arise, are analogous to the critical lines of 
phase transitions in statistical physics. The amplitude of bubbles
and waiting times between them respectively diverge with the critical 
exponents $\gamma = 1$ and $\nu = 1/2$, as the critical lines are approached.

\end{abstract}

\vskip 0.5cm

{\parindent=0pt
{\bf PACS}: 89.65.-s; 89.65.Gh; 89.75.Fb

\vskip1cm
{\bf *Corresponding author}: V.I. Yukalov

\vskip 2mm
{\bf E-mail}: yukalov@theor.jinr.ru

{\bf Phone}: +7 (496) 21 63 947

}

\newpage

\section{Introduction}

Much has changed since the last 10 years, when the so-called ``great moderation''
was regarded as the biggest success of modern economic theory, and research had
found that the term ``bubble'' was not an attractive explanation for the lack
of quantitative understanding of real estate prices \cite{Leung2004}. Now we
live in the ``great experiment'' era of central banks in which a plethora of
``code red'' policies \cite{MauldinCodeRed13}, such as quantitative easing,
zero interest rates, large-scale asset purchases (leading to de facto currency
``wars'' and their potential debasement) are being put forward, one after
another, without any certainty about the social and economic consequences that
these policies might bring about. The ``great experiment'' will likely continue
for one, perhaps two or even more decades (in different forms) as in the
Japanese scenario since 1990 and the bursts of their great market and real-estate
bubbles. Many new bubbles in part fueled by these policies are likely to emerge.

In this context, the present article investigates the impact of feedback loops
between asset prices and bond prices (or between monetary base and interest rate
spreads) that lead to regimes of periodically explosive bubbles followed by
crashes. While asset and bond prices tend to mean revert to their respective
fundamental prices, we propose a simple framework to account for the fact that
the latter are not fixed but depend themselves on the state of the economy
represented by the instantaneous asset and bond prices. This results in a rich
nonlinear coupled dynamics of asset and bond prices that we explore. Our work
is related to a rich literature on bubbles and crashes, which has documented
that, surprisingly from the point of view of the efficient market hypothesis,
they appear repeatedly with remarkable regularity \cite{Sornette_1}. 

There exists a number of models advanced to rationalize the appearance of these 
events. A common explanation to these episodes is to regard them, following
Keynes \cite{Keynes_2}, as {\it outbursts of irrational speculations}, such
as manias, panics, and similar herding and crowd behavior
\cite{Kindleberger_3,Malkiel_4,Galbraith_5}. Such a collective behavior
can be interpreted as leading to phase transitions, the bubbles bursting
in crashes, in an ensemble of interacting agents
\cite{Lux_6,Sornette_7,Kaizoji_8,Sornette_25}. 

Another approach is based on {\it multiple-equilibria models} 
\cite{Krugman_9,Ozdenoren_10,Ganguli_11,Cetin_12}, where several equilibria 
arise due to the existence of different types of traders, including dynamic 
hedgers. Then random motion can lead to jumps between different equilibria, 
which imitates crashes. Similar jumps can occur in markets with inexperienced 
traders \cite{Porter_13}. In these models, in the absence of dynamic hedgers 
or inexperienced traders, no bubbles and crashes occur. The nonlinear 
dynamical model of a market with speculative traders \cite{Yukalov_14} can 
also exhibit several equilibria, with jumps between different regimes caused 
by random noise. 

In {\it liquidity shortage models} \cite{Huang_15}, a crash can occur when 
price plummets due to a temporary occasional reduction in liquidity. 
{\it Bursting bubble models} assume a scenario when all market traders realize 
that an asset price is larger than its fundamental value, but they keep buying 
the asset in the belief there are others who do not yet know that the asset 
is overpriced, and to whom they expect to sell the asset at a higher price. 
At some random point of time, everyone (or a sufficient large fraction of 
investors) realizes that too many are aware of the overpricing, which results 
in the bubble burst and a crash 
\cite{Abreu_16,Scheinkman_17,Allen_18,Friedman_19,Brunnermeier_20}.

In {\it lumpy information aggregation models}, the existence of overpricing 
is known only to a fraction of the traders, while others are unsuspecting. 
At some point, the uninformed traders suddenly discover the existence of
overpricing, which leads to a sharp decline in prices imitating a crash
\cite{Romer_21,Caplin_22,Hong_23}. A sudden change in information can also
produce a jump in prices \cite{Zeira_24}. 

These different models thus explain booms and crashes as caused either by 
irrational speculations, or by sudden variations of some market 
characteristics, such as asymmetric or subjective information.

In the present article, we abstract from the specific micro-mechanism
to emphasize the role of delays and of nonlinear feedbacks \cite{SCFB15}.
We propose a reduced form set of two coupled continuous time equations
linking the price of a representative asset and the price of a bond, the 
latter quantifying the cost of borrowing. We argue that the presence of 
simple feedbacks between asset prices and cost of borrowing can induce, 
in a completely objective deterministic way, the appearance of periodically
exploding and collapsing bubbles ending in crashes. This mechanism does not
require heterogeneous agents, or sudden changes in system characteristics,
or multiple equilibria.

The plan of the article is as follows. In Sec. 2, we formulate the dynamical
model characterizing the relation between an asset price and a bond price.
In Sec. 3, we give the detailed analysis for the existence of stationary
solutions and their bifurcations. The various temporal behaviors, exhibiting
bubbles and crashes, are studied in Sec. 4. And Sec. 5 concludes.

\section{Model formulation}

We consider an economy represented by one representative asset and one bond
quantifying the cost of financing. Our goal is to derive simple dynamical
equations governing the interplay between asset and bond prices, which
emphasize the existence of highly nonlinear unstable regimes associated with
booms and crises. The interdependence between stocks and bonds is one of the
most important factors to asset allocation decisions. It also reflects the
impact of central banks monetary policies on the growth of economies. For
instance, Knut Wicksell pointed out that the greater the difference between
the `natural' interest rate and the `market' rate, the bigger the subsequent
booms and bursts \cite{Wicksell1898}.

A unified framework for pricing consistently all assets is expressed in terms
of a stochastic discount factor (SDF) $M(t)$, also called pricing kernel, 
pricing operator, or state price density \cite{Cochrane01}. In this framework, 
the price of an asset is equal to the sum of expected future payoffs discounted
with the SDF, which embodies all the macro-economic risks underlying each
security's value. Under an adequate definition of the space of admissible
trading strategies, the no-arbitrage condition translates into the condition
that the product of the SDF with the asset price $p(t)$ of any admissible
self-financing trading strategy, implemented by trading on a financial asset,
must be a martingale,
\be
M(t) p(t) = {\rm E}[M(t') p(t')]~,
\label{tyjiujh}
\ee
where $t'>t$ refers to a future date and $t$ is the present time. The
expectation ${\rm E} [.]$ is taken with respect to all the available 
information up to time $t$. Introducing the discount factor from $t$ 
to $t'$ defined by $m(t,t') : = M(t')/[M(t)$, expression (\ref{tyjiujh}) 
becomes
\be
p(t) = {\rm E}[m(t,t') p(t')]~.
\label{stjmiurd}
\ee
The discount factor $m(t,t')$ also characterizes  the cost of financing
\cite{Gollier_26}.

Let us consider the zero-coupon bond $B(t, t+\tau)$ that matures at time $\tau$
in the future. We fix $\tau$ to represent the typical investment horizon of
a representing investor, say three or six months. By definition,
$B(t+\tau, t+\tau) = 1$ and, thus, because of Eq. (\ref{stjmiurd}), one has
\be
B(t, t+\tau) = {\rm E}[m(t,t+\tau)]~.
\label{stjmiurfeqd}
\ee
The bond price $B(t, t+\tau)$ thus embodies the information on the discount
factor and the price of financing in the economy.

We assume that there is a natural equilibrium price $p^*$ for the asset and
$B^*$ for the bond value considered above. Close to this equilibrium, we
assume that $p(t)$ and $B(t, t+\tau)$ obey the simplest two-side exponential
mean reversal dynamics towards the equilibrium:
\be
\label{1}
\frac{dp}{dt} = p - \; \frac{p^2}{p^*} \; , \qquad
\frac{dB}{dt} = B - \; \frac{B^2}{B^*} \; ,
\ee
where we call $B(t) := B(t, t+\tau)$ for simplicity of notations. The constant
coefficients in the right-hand-side of these two equations have been absorbed 
in a rescaling of time and in the definition of $p^*$ and $B^*$. These equations
are known as logistic equations, and enjoy the following properties: (i) $p(t)$
and $B(t)$ converge, respectively, to $p^*$ and $B^*$; (ii) the convergence is 
exponentially fast, whatever the initial values, smaller or larger than the
fixed points $p^*$ and $B^*$. This represents a convenient concise description
of a mean-reversal dynamics.

In reality, $p^*$ and $B^*$ are not fixed but depend themselves on the specific
state of the economy, which we can characterize by the instantaneous value $p(t)$
and $B(t)$. Intuitively, as the bond price increases (interest rates decrease),
the price level tends to grow due to cheap financing. As the price increases,
this tends to push later the rate upward as access to funding competes with
investments in the appreciating asset and as central banks attempt to cool
down a possible over-heating of the economy. As the price accelerates, the bond
price may decrease drastically, the interest rates shoot up, and then a price
crash may ensue. To capture this intuition, and given the structure of
expressions (\ref{stjmiurd}) and (\ref{stjmiurfeqd}), we assume that $p^*$ is
a function of the product $\pi(t): = p(t)B(t)$, while $B^*$ is just a function
of $p(t)$.

The explicit dependence of $p^*$ on $\pi(t)$ and of $B^*$ on $p(t)$ can be
derived in two ways producing the same result. The simplest way is to assume
that the relative variation of $p^*$ is proportional to the variation of $\pi(t)$:
\be
\dlt p^* \propto p^* \dlt \pi(t) ~.
\label{tukyhgfw}
\ee
Similarly, the relative variation of $B^*$ is proportional to the variation
of $p(t)$:
\be
\dlt B^* \propto B^* \dlt p(t) ~.
\label{hgdvaa}
\ee
This yields
\be
\label{2}
 p^*(\pi) = p^*(0) e^{b\pi} \; ~~~~~~ B^*(p) = B^*(0) e^{gp}\;  ,
\ee
where the parameter $b$ is a fundamental log-price rate and $g$ is a 
log-discount rate:
\be
b = \frac{\prt \ln  p^*}{\prt \pi} \; , \qquad
g = \frac{\prt\ln  B^*}{\prt p} \;   .
\ee
Without loss of generality, it is possible to reduce the evolution equations
(\ref{1}) with (\ref{2}) to dimensionless normal forms
\cite{Yukalov_27,Yukalov_28,Yukalov_29}. Denoting
\be
x(t) := p(t)~,~~~~~~~z(t) : = B(t)~,
\ee
we obtain the equations
\be
\label{5}
\frac{dx}{dt} = x -  x^2 e^{-bxz} \; , \qquad
\frac{dz}{dt} = z - z^2 e^{-gx} \;   .
\ee
In what follows, we keep in mind this reduction and that all quantities are
dimensionless. By definition, the asset and bond prices are non-negative
$x \geq 0$ and $z \geq 0$.

The structure of the coupled nonlinear logistic equations (\ref{5}) is
reminiscent of the dynamics of symbiotic biological species with population
fractions $x$ and $z$, influencing each other through their carrying capacities 
\cite{Yukalov_27,Yukalov_28,Yukalov_29}, or of the predator-prey
dynamics with mutual interactions between predators and preys \cite{Arditi_30}.
Another analogy is the co-evolution of different groups in a structured society
\cite{Perc_54,Perc_55}. However, the basic point in our evolution equations is 
that the functions $x$ and $z$ are not different species, but the asset and 
bond prices in a market. 

When interest rates decrease, the effective bond price increases, increasing
the effective price $xz$, and the fundamental price level $p^*$ grows. Then,
since $p^* \sim e^{bxz}$, this implies that the rate $b$ has to be positive
or, more generally, non-negative to represent a realistic economic regime.
In turn, the increase of the asset price $x$ tends to trigger eventually an
increase in the cost of investments, which tends to push down the effective
bond prices, which means that the coefficient $g$ is usually negative
(non-positive). Thus, the standard situation in a financial market corresponds
to the conditions
\be
\label{3}
 b \geq 0 \; , \qquad g \leq 0 \;  .
\ee

The other way of deriving the equations (\ref{5}) is to assume that
$\pi(t)$ and $p(t)$ are initially small so that $p^*$ and $B^*$ can be
represented in the form of Taylor expansions
\be
p^* =  \sum_{n=0}^\infty \frac{c_{n}}{n!} \; \pi^n~,~~~~  B^* =
\sum_{n=0}^\infty \frac{d_{n}}{n!} \; p^n ~.
\ee
Then, it is possible to extrapolate the above expansion to finite values
of $\pi(t)$ and $p(t)$ by invoking resummation techniques, for instance
self-similar approximation theory \cite{Yukalov_31,Yukalov_32,Yukalov_33}.
In the course of the resummation, it is necessary to impose the restriction
of non-negativity of $p^*(t)$ and $B^*(t)$ \cite{Saavedra_34}. As a result
of such a self-similar resummation, under the condition of semi-positivity
of $p^*$ and $B^*$, we come to the same form (\ref{2}) corresponding to a
self-similar exponential approximant \cite{Yukalov_35}.

The above derivation of the coupled dynamics (\ref{5}) of asset and bond
prices does not include any stochastic component as we consider averaged
macro-quantities in order to emphasize the nonlinear feedback loops between
asset and bond prices. In reality, a stochastic structure would need to be
added, which can be both additive and multiplicative (corresponding to
parametric noise). Here, we focus on the deterministic structure of the
dynamics to unravel the main consequences of the feedback loops. In 
particular, we uncover a regime of periodically collapsing bubbles with 
remarkable properties.

\section{Stationary solutions of evolution equations}

Equations (\ref{5}) possess the following trivial fixed points: $\{0,0\}$, 
$\{1,0\}$, and $\{0,1\}$. In addition, there are nontrivial fixed points, 
which are the solutions of the equations
\be
\label{6}
x^* = \exp( bx^* z^*) \; , \qquad z^* = \exp(gx^*)  ~ .
\ee
The stability of each fixed point is described by the characteristic exponents
\be
\label{7}
\lbd_{1,2} = \frac{1}{2} \; \left [
bx^* z^* - 2 \pm x^* \sqrt{bz^*(4g+bz^*)} \right ] \;   .
\ee
A fixed point is stable (respectively, unstable) for negative (respectively, 
positive) real parts of the exponents. The trivial fixed points $\{0,0\}$, 
$\{1,0\}$, and $\{0,1\}$ are always unstable for all values of $b$ and $g$.

Depending on the parameters $b$ and $g$, up to three nontrivial fixed points 
$\{x_1^*,z_1^*\}$, $\{x_2^*,z_2^*\}$ and $\{x_3^*,z_3^*\}$ can exist, which
we enumerate by descending values $x_1^* \geq x_2^* \geq x_3^*$ of the
reduced asset price variable. According to condition (\ref{3}), we consider
$b \geq 0$. Although the parameter $g$ should be non-positive, we also study
the regime with $g$ positive in order to exemplify the analytic continuation
of the fixed points through the boundary $g = 0$. This analytic continuation
allows us to explain the dynamic behavior of the solutions when approaching
this boundary $g = 0$.

The $b$-$g$ plane is partitioned into four regions by the lines $g = 0$,
$g = g_0(b)$, and $g = g_c(b)$, as is shown in Fig. 1. The four regions are
as follows.

{\bf A}. When either
\be
\label{8}
 0 < b < b_0 \; , \qquad g < g_0(b) \;  ,
\ee
or
\be
\label{9}
 b > b_0 \; , \qquad g < g_c(b) \;  ,
\ee
there exists a single fixed point $\{x_3^*,z_3^*\}$ that is a stable focus,
which degenerates to a stable node in the vicinity of the line $g_0(b)$.

{\bf B}. When either
\be
\label{10}
0 < b < \frac{1}{e} \; , \qquad g_0(b) < g < 0 \;   ,
\ee
or
\be
\label{11}
 \frac{1}{e} < b < b_0 \; , \qquad g_0(b) < g < g_c(b) \;  ,
\ee
there are three fixed points, an unstable focus $\{x_1^*,z_1^*\}$, a saddle
$\{x_2^*,z_2^*\}$, and a stable node $\{x_3^*,z_3^*\}$.

{\bf C}. For
\be
\label{12}
b > \frac{1}{e} \; , \qquad g_c(b) < g < 0 \;   ,
\ee
there is an unstable focus $\{x_1^*,z_1^*\}$ and there appears a limit cycle
around this focus.

{\bf D}. When
\be
\label{13}
 0 < b < \frac{1}{e} \; , \qquad 0 \leq g < g_c(b)  \;   ,
\ee
there are two fixed points, a saddle $\{x_2^*,z_2^*\}$ and a stable node
$\{x_3^*,z_3^*\}$.

{\bf E}. Finally, when either
\be
\label{14}
 0 < b < \frac{1}{e} \; , \qquad  g > g_c(b)  \;   ,
\ee
or
\be
\label{15}
  b > \frac{1}{e} \; , \qquad  g \geq 0 \; ,
\ee
there are no fixed points.

The transformation of fixed points, when moving from one region to another,
can be illustrated by the corresponding bifurcation paths.

\vskip 2mm

${\bf A \longrightarrow B \longrightarrow D \longrightarrow E}$: this 
bifurcation path is shown in Fig. 2, when moving from region $A$, through 
$B$ and $D$, to region $E$ along the line $b = 0.2$, for which $0 < b < 1/e$. 
Then $g_0(b = 0.2) = - 0.0194$ and $g_c(b = 0.2) = 0.276$. When the stable 
focus $\{x_3^*,z_3^*\}$ in the region $A$ approaches $g_0(b)$, it degenerates 
to a stable node at $g_n = - 0.0471$ and then survives in the region $B$ where,
in addition, there appear an unstable focus $\{x_1^*,z_1^*\}$ and a saddle
$\{x_2^*,z_2^*\}$. Moving further from the region $B$ to the region $D$,
there remain only the stable node $\{x_3^*,z_3^*\}$ and the saddle
$\{x_2^*,z_2^*\}$. Going from region $D$ to region $E$ by crossing $g_c(b)$,
all fixed points cease to exist.

\vskip 2mm

${\bf A \longrightarrow B \longrightarrow C \longrightarrow E}$: this 
bifurcation path along the line $b = 0.4$, such that $ 1/e < b < b_0 = 0.47$, 
is illustrated in Fig. 3. Here $g_0(b=0.4)=-0.0552$ and $g_c(b=0.4)=-0.0294$. 
Moving upward in the region $A$, the stable focus $\{x_3^*,z_3^*\}$ becomes 
a stable node at $g_n = - 0.0849$. The stable node $\{x_3^*,z_3^*\}$ survives
into region $B$, where there appear two more fixed points, an unstable focus
$\{x_1^*,z_1^*\}$ and a saddle $\{x_2^*,z_2^*\}$. In region $C$, the unstable
focus $\{x_1^*,z_1^*\}$ remains and there appears a limit cycle around it. 
In the region $E$, where $g > 0$, there are no fixed points.

\vskip 2mm

${\bf A \longrightarrow C \longrightarrow E}$: the bifurcation path along the
line $b = 1 > b_0 = 0.47$ is given in Fig. 4. Then $g_c(b = 1) = - 0.1769$. The
stable focus $\{x_3^*,z_3^*\}$ in region $A$ transforms in region $C$ into
an unstable focus $\{x_1^*,z_1^*\}$ and there appears a limit cycle. The
characteristic exponents at the critical line $g_c(b)$ are purely imaginary.
No fixed points exist in the region $E$.

\vskip 2mm

${\bf A \longrightarrow B \longrightarrow C}$: the bifurcation path along the
line $g = - 0.03$ is presented in Fig. 5. The critical lines are crossed at
the points $b_1 \equiv b_0(g = - 0.03) = g_0^{-1}(g = - 0.03) = 0.2718$ and
$b_2 \equiv b_c(g = - 0.03) = g_c^{-1}(g = - 0.03) = 0.4007$. The stable focus
$\{x_3^*,z_3^*\}$ in region $A$ becomes a stable node at point $b_n = 0.1242$
that survives into region $B$, where an unstable focus $\{x_1^*,z_1^*\}$ and
a saddle $\{x_2^*,z_2^*\}$ appear. At the boundary $b_2$, the stable node
$\{x_3^*,z_3^*\}$ and a saddle $\{x_2^*,z_2^*\}$ merge together, while the
unstable focus $\{x_1^*,z_1^*\}$ continues to the region $C$, where a limit
cycle arises.

\vskip 2mm

${\bf A \longrightarrow C}$: the bifurcation path along the line $g = - 0.2$
is shown in Fig. 6. The critical line $g_c(b)$ is crossed at the point
$b_2 \equiv b_c(g = -0.2) = g_c^{-1}(g = -0.2) = 1.1864$. Here, the stable
focus $\{x_3^*,z_3^*\}$ in region $A$ transforms into an unstable focus
$\{x_1^*,z_1^*\}$ in region $C$, where also a limit cycle appears.

\vskip 2mm

The limit cycle, representing periodically occurring bubbles and crashes in
asset prices, appears in two cases. One is the path $A \longrightarrow C$,
with crossing the critical line $g_c(b)$, accompanied by the bifurcation
{\it stable focus} $\Longrightarrow$ {\it unstable focus} $+$ {\it limit cycle},
which is a typical Hopf bifurcation.

The second case happens when crossing the line $g_c(b)$ in the path
$B \longrightarrow C$, with the bifurcation {\it stable node} $+$ {\it saddle}
$+$ {\it unstable focus} $\Longrightarrow$ {\it unstable focus} $+$
{\it limit cycle}.

\section{Periodically collapsing bubbles}

\subsection{Different types of bubbles}

We find periodically sharply increasing and fast decreasing prices in 
region $C$. The amplitude, periodicity, as well as the shape of the price 
trajectories depend on the parameters $b$ and $g$. Figure 7 shows the 
behavior of the asset price $x(t)$ and bond prices z(t), where $x(t)$ and 
$z(t)$ are out of phase: an increase in the asset price is accompanied by 
a decrease in the bond price and vice-versa. In this parameter regime, the 
price rise and decay are symmetric, representing a typical business cycle 
regime. In contrast,  Fig. 8 shows a parametrization, for which the price 
of the asset is exhibiting a bubble-like trajectory followed by a sharp 
faster correction. The prices remain periodic and the structure is that 
of periodically collapsing bubbles. We find that the asymmetry increases 
when $g$ approaches the line $g = 0$, as is demonstrated in Fig. 9. The 
distance between the bubbles increases when approaching the line $g_c(b)$ 
with $1/e < b < b_0$, as illustrated in Figs. 10 and 11.

We find that the maximum of the asset price $x(t_{max})$ occurs 
at a time $t_{max}$ that always precedes the time $t_{min}^z$, where the 
bond price $z(t_{min}^z)$ has its minimum. The time lag 
$\Delta t \equiv t_{min}^z - t_{max}$ depends on the parameters $b$ and $g$.
Rather than considering the absolute value of $\Delta t$, we quantify
its relative value reduced to the bubble width $w$ measured at 
the half amplitude of the corresponding bubble. Varying the parameters $b$
and $g$ inside the region $C$, we find that the relative asset-bond time 
lag $(\Delta t / w) \times 100 \%$ is of order $10 \%$. 

An economic interpretation is as follows.  The price rise during a bubble
reflects an exuberant market that anticipates a booming economy. 
Two mechanisms concur to drive the bond
prices lower, and thus the interest rate higher. As the economy is booming,
it needs investment for growth. There is thus a scarcity of money, which
makes its cost larger. In other words, the cost of credit increases and the
interest rates rise. At the macro level, the central bank is also often tempted
to moderate the exuberance of the bubble by making more expensive the
access to credit, thus increasing the short-term interest rate that it can control.
The fact, that the maximum of the asset price $x(t_{max})$ occurs 
at a time $t_{max}$ that always precedes the time $t_{min}^z$, where the 
bond price $z(t_{min}^z)$ has its minimum, means that the credit dynamics
and/or central bank intervention lags behind the price dynamics.
This is in agreement with detailed analyses that have shown that stock markets in the US
in general lead the Fed rate interventions  \cite{Zhousorslav04,GuoZhousorslav14}.
If we consider that the dotcom bubble developed from Jan. 1997 to March 2000, covering roughly 
39 months, the relative lag of 10\% we find in our model translates into about 4 months
calendar time, which is in agreement with the estimation of the lag of about a quarter
\cite{Zhousorslav04,GuoZhousorslav14}.

\subsection{Least stable phase during bubbles}

As seen in Figs. 8-11, the rising part of a bubble phase is characterized by
a fast acceleration, which will be quantified precisely below. Given this
strong growth regime, one can expect the prices to be more and more susceptible
to external influences and shocks. Does the maximum susceptibility occur
during the bubble ascent, at its maximum, or later during the crash?

In the case of a strong asymmetry between the bubble ascent phase and the
sharp crash correction, it is tempting to use an analogy with phase transitions
and critical phenomena, where the point of phase transitions would correspond
to the asset price peak. Then, one could expect that the response functions
\cite{Sieber_36} would be maximal at the points of phase transitions,
where the system is supposed to be unstable \cite{Yukalov_37,Sornette_38}.
But our dynamical system is globally stable as a result of the feedback loops
between the asset and bond prices. This translates into a globally stable
periodic attractor, notwithstanding the very nonlinear nature of the oscillations,
which we have termed ``periodically collapsing bubbles'' on reference to their
shapes. Thus, the correct way to quantify the susceptibility is via a local
stability measure called the relative expansion exponent 
\cite{Yukalov_39,Yukalov_40}, defined as the sum of the local Lyapunov exponents: 
the larger the expansion exponent, the lower the system stability.

For the dynamical system (\ref{5}), the local expansion exponent is defined as
\cite{Yukalov_39,Yukalov_40}
\be
\label{16}
 \Lbd \equiv \frac{1}{t} \int_0^t {\rm Tr} \hat J(x(t'),z(t') ) \; dt' \;  ,
\ee
where $\hat{J}(x,z)$ is the Jacobian matrix
\begin{eqnarray}
\label{17}
\hat J(x,z) = \left [ \begin{array}{cc}
\frac{\prt f_1}{\prt x } & ~~ \frac{\prt f_1}{\prt z} \\
\\
\frac{\prt f_2}{\prt x } & ~~ \frac{\prt f_2}{\prt z}
\end{array}
\right ]
\end{eqnarray}
for the dynamical system (\ref{5}) with
$$ 
f_1 := x -  x^2 e^{-bxz} \; , f_2 := z - z^2 e^{-gx} \; .
$$
Numerical  investigation shows that the expansion exponent $\Lambda$ is always
negative, which confirms that the dynamical system is always stable in the sense
that the dynamics remains bounded and non-chaotic. The largest value of
$\Lambda$ corresponds to the least stable state. It turns out that the market,
described by the system of Eqs. (\ref{5}), is the least stable not where the asset
price reaches its maximal value. This is illustrated in Fig. 12 for the same
parameters ($b = 0.4$ and $g = - 0.029$) used in Fig. 11. The asset price reaches
its first maximum $x(t_{max}) = 61.717$ at $t_{max} = 122.89$, where the expansion
exponent presents a local minimum $\Lambda(t_{max}) = - 0.878$. The algebraic
maximum of the expansion exponent $\Lambda(t_\Lambda) = - 0.864$ occurs at the
time $t_\Lambda = 121.48$, which is slightly earlier than $t_{max}$. Thus, the
asset price is the least stable not at its maximum but just before it.

\subsection{Quantification of exploding bubbles}

We have found that the asset price $x(t)$ can be well approximated during
the bubble explosion by the following analytical form
\be
\label{18}
 x_{app}(t) = \frac{c_1}{(t_\Lbd-t)^\bt} \; \exp \left \{
\frac{c_2}{(t_\Lbd-t)^\al } \right \} \;  ,
\ee
where $t_\Lambda$ is the point of maximum of the expansion exponent
$\Lambda(t)$.

Our numerical investigations lead us to conclude that the exponents
$\alpha$ and $\beta$, for the values of $b$ and $g$ in the left corner
of the region $C$, between region $B$ and $E$, where $1/e < b < b_0$ and
$g > g_c(b)$, are close to
\be
 \al = \frac{2}{5} \; ,  \qquad \bt = \frac{1}{5} ~.
 \label{miigw;b;}
 \ee
The numerical coefficients  $c_1$ and $c_2$ are positive parameters that
depend on the specific values of $b$ and $g$. For instance, for
$b = 0.4$ and $g = - 0.029$, we have  $c_1 = 2.8$, $c_2 = 1.5$, and
$t_\Lambda = 121.48$. 

It is interesting to find out whether there exists a critical region where 
the parameters $\alpha$ and $\beta$ are invariant, being close to 
$\alpha = 0.4$ and $\beta = 0.2$, respectively. We have accomplished a 
detailed analysis over the whole region $C$. It turned out that 
approximation (\ref{18}) describes very well the transitional behavior of
the beginning of the bubble bursting for the parameters $b$ and $g$ in the 
vicinity of the double critical point $\{b = 1/e, \; g = 0\}$, which is the 
intersection point of the critical lines $b = b_c(g)$ and $g = 0$, where 
the exponents $\alpha$ and $\beta$ play the role of a kind of critical 
indices. 

Moreover, the super-exponential form of the initial stage of bursting 
bubbles is well represented by approximation (\ref{18}) in a finite region 
of the parameters $b$ and $g$ corresponding to the triangle, where    
$1/e < b < b_0$ and $g > g_c(b)$. In this critical region, the exponents 
$\alpha$ and $\beta$ are invariant, and only the parameters $c_1$ and $c_2$
have to be adjusted. For example, let us take $b = 0.38$ and $g = -0.0117$,
for which $g_c = -0.01174$. Then the asset price reaches its first maximum 
$x(t_{max}) = 393.5$ at $t_{max} \approx 501.5$, where the expansion exponent 
possesses its local minimum $\Lambda(t_{max}) = -0.9607$, while the absolute 
maximum of the expansion exponent $\Lambda(t_\Lambda) = -0.9535$ occurs at 
the time $t_\Lambda \approx 499.74 < t_{max}$. Then the critical exponents 
$\al = 0.4$ and $\bt = 0.2$ can be kept invariant, and only the parameters 
$c_1\approx 1.5$ and $c_2\approx 2.8$ are varied. 

In order to give a deeper understanding justifying why and where the 
super-exponential form (\ref{18}) serves as a good description of the 
transition from a smooth, almost unchangeable, behavior of the asset price 
to the start of a bursting bubble, let us study the ratio $R = width/period$ 
defined as the ratio of the temporal bubble width, measured at half of its 
amplitude, to the temporal period between the bubbles. It is reasonable to 
expect that, for genuine bubbles, the ratio $R$ has to be small, representing 
sharp bubble bursts and crashes interspersed within a smooth market behavior. 
Our numerical investigation shows that the super-exponential form (\ref{18}),
 with invariant exponents $\al = 0.4$ and $\bt = 0.2$, is valid as long as 
$R \ll 1$, which occurs in the triangle region where $1/e<b< b_0\approx 0.47$ 
and $g_c < g < 0$.  

The super-exponential behavior (\ref{18}) is reminiscent of the many
reported empirical cases of financial bubbles
\cite{Sornette_1,Yanetal09,JohSornette10,Jiangetal10,Husleretal13,Sornetteetal13},
whose structure can serve as a precursor of the appearing bubble \cite{SCFB15}.
We note that the transient super-exponential explosive bubble trajectory is here
stronger than the simple power law singularities discussed elsewhere
\cite{Jiangetal10,Husleretal13,Sornetteetal13} as form (\ref{18}) represents 
an intermediate asymptotic with an essential singularity behavior. Figure 13
demonstrates the excellent accuracy of the approximation (\ref{18}), as compared 
to the accurate ``exact'' numerical solution $x(t)$ in the region of its 
accelerated growth.

\subsection{Period and amplitude of the periodically collapsing bubbles}

The shape of the bubbles and the distance between them strongly depends on
the closeness of the parameters to the critical lines $g = g_c(b)$ and
$g = 0$. This results from the existence of a stable node in the region $B$,
which influences the length of the plateaus between the bubbles in the region $C$,
as is seen in Figs. 8 and 11. Increasing the distance from the boundary,
separating the regions $B$ and $C$, that is, increasing $b$ with fixed $g$,
shortens the plateau length between two successive bubbles, but does not
influence much the bubble amplitude. This is illustrated in Fig. 14. The level
of the plateaus does not depend on initial conditions, as is demonstrated
in Fig. 15.

The boundary $g_c(b)$ plays the role of a critical line at which the plateau
length diverges, similarly to the divergence of the correlation length of
statistical physics systems at a phase transition line \cite{Yukalov_37,Sornette_38}.
To describe this divergence, let us define the duration
\be
\label{19}
L \equiv t_{n+1}-t_n
\ee
between two successive maxima of the asset price $x(t_n)$ occurring at the
times $t_{n}$ and $t_{n+1}$. In other words, $L$ is the time
interval between two successive bubbles. This interval, for a fixed $g$,
essentially depends on the deviation
\be
\label{20}
\Dlt \equiv b - b_c(g)
\ee
of $b$ from the boundary point $b_c(g) \equiv g_c^{-1}(g)$. The critical
behavior can be characterized by the power law
\be
\label{21}
 L = L(\Dlt) \propto \Dlt^{-\nu} \qquad
(\Dlt \ra +0 ) \;  .
\ee
Hence, the critical index is given by the limit
\be
\label{22}
 \nu = - \lim_{\Dlt\ra+0} \; \frac{d\ln L}{d\ln|\Dlt|} \;  .
\ee
Figure 16a shows the behavior of $\ln L$ as a function of $\Delta$, from
where we determine $\nu = 0.5$.

The other critical line is $g = 0$ at which the bubble amplitudes diverge.
Above this line, the behavior of the asset price is exponentially inflating.
Approaching the critical line $g = 0$ from below, the bubble amplitude
$A \equiv x(t_{max})$ diverges, analogously to the behavior of
the compressibility in statistical physics systems close to a phase
transition line, as
\be
\label{24}
 A = A(g) \propto g^{-\gm} \qquad (g\ra 0^- ) \;  .
\ee
This defines the critical index
\be
\label{25}
 \gm = -\lim_{g\ra -0} \; \frac{d\ln A}{d\ln|g|} \;  .
\ee
The dependence of $\ln A$ versus $\ln |g|$ is shown in Fig. 16b, from which
we determine $\gamma = 1$. Table 1 illustrates the fast divergence of the
bubble amplitude $A$, when decreasing $g$ with fixed $b = 1$, the number $N$
of bubbles in the time interval $[0,100]$, and the bubble width $w$ measured
at half-amplitude. The bubble width, as $g \ra -0$, saturates to $0.7$, while
its amplitude diverges.

\section{Concluding remarks}

We have identified a specific structure of feedback loops between asset prices
and bond prices that lead to the regimes of periodically explosive bubbles
followed by crashes. While asset and bond prices tend to mean revert to their
respective fundamental prices, we have argued that the later are not fixed
but depend themselves on the state of the economy represented by the
instantaneous asset and bond prices. This results in a rich nonlinear coupled
dynamics of asset and bond prices. The two key parameters $b$ and $g$ represent 
respectively the sensitivity of the fundamental asset price on past asset and
bond prices and of the fundamental bond price on past asset prices. They thus
capture the competition of the allocation of capital between assets and bonds
that develops as the two exhibit diverging trajectories. We stress that the
regime of periodically collapsing bubbles does not require heterogeneous agents
or sudden changes in market characteristics. Periodically occurring bubbles
and crashes can be treated as natural phenomena objectively arising in stock
markets due to the delay of the impact of present valuations on future
fundamental prices and the nonlinear couplings.

The coupled nonlinear equations (\ref{5}) have been motivated as the 
description of the dynamics of a representative asset price coupled with a 
bond price. There are other possible interpretations. For instance, we can 
think of the variable $x(t)$ as the monetary base of an economy and $z(t)$ 
as the spread (or difference) between the yield (interest rate) of bonds with 
long term maturity (say 10 years) and the interest rate of bonds with short 
term maturity (say 3 months). The spread $z(t)$ provides a measure of the 
market expectation on the growth potential of the economy. The larger $z(t)$, 
the larger the expected growth rate. When $x$ is increasing and $z$ is large, 
this means that short-term interest rates are low (cheap funding and borrowing) 
and long term interest rates are increasing (expectation of a growing economy 
with growing returns). This boosts the monetary base on the anticipation of 
a growing economy and as a result of the growth of the economy that needs 
financing. This is the mechanism by which a larger $z$ promotes a future larger 
monetary base $x$. The multiplicative nature of the equations (\ref{5}) capture 
the fact that the positive effect of $z$ on $x$ is all the larger, the larger 
are $x$ and $z$. But as the monetary base $x$ increases, it generally grows 
faster than the real economy, leading to misallocation of funding to non-performing 
industry sectors. This then tends to push $z$ down, on the expectation that 
long-term interest rates will become smaller as a result of the slowing down 
of the economy. Also, the economy in its boom phase is over heating and usually 
requires the intervention of the monetary authorities, who are led to increase 
the short term interest rates (the cost of short-term borrowing) to stabilize 
the economy. The larger $x$, the more negative its impact on $z$, hence on its 
limiting fundamental value. This is captured by the dynamic equations (\ref{5}). 
Thus, the regimes of parameters in region $C$, that we have reported in this 
article, demonstrate the difficulties in stabilizing an economy with the usual 
monetary tools consisting in attempt to control the monetary base and the 
interest rates. Due to intrinsic delay and nonlinear feedbacks, such attempts 
can lead to unintended cycles of booms and busts, as documented for the period 
from 1980 to present and its role in the great crisis of 2008, following great 
recession and on-going global crises \cite{SC19802015}.

\newpage

\begin{center}
{\bf {\Large Figure captions} }
\end{center}

\vskip 1cm
{\bf Figure 1}. Region of existence of stationary solutions corresponding
to the fixed points of the evolution equations (\ref{5}) for the asset price $x$
and bond price $z$.

\vskip 5mm
{\bf Figure 2}. Bifurcation path
${\bf A \longrightarrow B \longrightarrow D \longrightarrow E}$.
Here $b = 0.2 < 1/e$ is fixed and $g$ varies: (a) asset price $x^*$;
(b) bond price $z^*$. The fixed point $\{x_3^*,z_3^*\}$ (solid line)
is a stable focus transforming into a stable node (solid line with dots)
at $g_n \approx -0.0471$. The point exists for $- \infty < g < g_c$.
The fixed point $\{x_2^*,z_2^*\}$ (dashed-dotted line) is a saddle,
which exists for $g_0 < g < g_c$, and the fixed point $\{x_1^*,z_1^*\}$
(dashed line) is an unstable focus, which exists for $g_0 < g < 0$.
At $g = g_0$, the fixed points $\{x_1^*,z_1^*\}$ and $\{x_2^*,z_2^*\}$
coincide. At $g = g_c$, the fixed points $\{x_3^*,z_3^*\}$ and
$\{x_2^*,z_2^*\}$ coincide. When $g \ra -0$, then $x_1^* \ra +\infty$
and $z_1^* \ra 0$. When $g \ra -\infty$, then $x_3^*\ra 1$ and $z_3^*\ra 0$.
When $g > g_c$, fixed points do not exist.

\vskip 5mm

{\bf Figure 3}. Bifurcation path
${\bf A \longrightarrow B \longrightarrow C \longrightarrow E}$.
Here $1/e < b = 0.4 < b_0$ and $g$ varies: (a) asset price $x^*$;
(b) bond price $z^*$. The fixed point $\{x_3^*,z_3^*\}$ (solid line)
is a stable focus transforming to a stable node (solid line with dots) at
$g_n \approx -0.0849$. The point exists for $- \infty < g < g_c < 0$.
The fixed point $\{x_2^*,z_2^*\}$ (dashed-dotted line) is a saddle,
which exists for $g_0 < g < g_c$, and the fixed point $\{x_1^*,z_1^*\}$
(dashed line) is an unstable focus, which exists for $g_0 < g < 0$.
At $g = g_0$, the fixed points $\{x_1^*,z_1^*\}$ and $\{x_2^*,z_2^*\}$
coincide. At $g = g_c$, the fixed points $\{x_3^*,z_3^*\}$ and
$\{x_2^*,z_2^*\}$ coincide. When $g \ra -0$, then $x_1^* \ra +\infty$
and $z_1^* \ra 0$. When $g \ra - \infty$, then $x_3^* \ra 1$ and
$z_3^* \ra 0$. When $g \geq 0$, fixed points do not exist.

\vskip 5mm

{\bf Figure 4}. Bifurcation path
${\bf A \longrightarrow C \longrightarrow E}$.
Here $b = 1 > b_0$ is fixed and $g$ varies: (a) asset price $x^*$;
(b) bond price $z^*$. The fixed point $\{x_3^*,z_3^*\}$ (solid line)
is a stable focus, and the fixed point $\{x_1^*,z_1^*\}$ (dashed line) is
an unstable focus. At $g = g_c \approx -0.1769$, the stable focus transforms
into an unstable focus with $x_3^* = x_1^* = e^2$, $z_3^* = z_1^* = 2/(be^2)$,
the Lyapunov exponents being ${\Re} \lbd_{1,2} = 0$.

\vskip 5mm

{\bf Figure 5}. Bifurcation path
${\bf A \longrightarrow B \longrightarrow C}$.
Here $g = -0.03$ is fixed and $b$ varies: (a) asset price $x^*$;
(b) bond price $z^*$. The fixed point $\{x_3^*,z_3^*\}$ (solid line)
is a stable focus transforming to a stable node (solid line with dots) at
$b_n \approx 0.1242$. The point exists for $0 \leq b < b_2$, where
$b_2 \approx 0.4007$. The fixed point $\{x_2^*,z_2^*\}$ (dashed-dotted line)
is a saddle, which exists for $b_1 < b < b_2$, where $b_1 \approx 0.2718$.
The fixed point $\{x_1^*,z_1^*\}$ (dashed line) is an unstable focus, which
exists for $b > b_1$. At $b = b_1$, the fixed points $\{x_1^*,z_1^*\}$ and
$\{x_2^*,z_2^*\}$ coincide. At $b = b_2$, the fixed points $\{x_3^*,z_3^*\}$
and $\{x_2^*,z_2^*\}$ coincide. When $b = 0$, then $x_3^* = 1$ and $z_3^* = e^g$.
When $b \ra +\infty$, then $x_1^* \ra +\infty$ and $z_1^* \ra 0$.

\vskip 5mm

{\bf Figure 6}. Bifurcation path ${\bf A \longrightarrow C}$. Variation
of fixed points under fixed $g = -0.2$ and changing $b$: (a) asset price $x^*$;
(b) bond price $z^*$. The fixed point $\{x_3^*,z_3^*\}$ (solid line) is
a stable focus that at $b = b_2 \approx 1.1864$ transforms into an unstable
focus $\{x_1^*,z_1^*\}$ (dashed line). At $b = b_2$, the fixed points
$\{x_3^*,z_3^*\}$ and $\{x_1^*,z_1^*\}$ coincide, with the Lyapunov exponents
being ${\Re}\lbd_{1,2} = 0$.

\vskip 5mm

{\bf Figure 7}. Behavior of the solutions for the asset price and bond price
for the parameters $b = 0.5$, $g = -0.083$, and the initial conditions $x_0 = 3$
and $z_0 = 0.1$. Here, $g_c \approx -0.083056$. (a) Asset price $x(t)$ for
$t \in [0,600]$; (b) bond price $z(t)$ for $t \in [0,600]$; (c) a single
zoomed out bubble of the asset price $x(t)$ for $t \in [495,510]$; (d) a zoomed
out negative bubble of the bond price $z(t)$ for $t \in [495,510]$.

\vskip 5mm

{\bf Figure 8}. Asset price and bond price for the parameters $b = 0.4$,
$g = -0.0294$, and the initial conditions $x_0 = 3$ and $z_0 = 0.1$. Here,
$g_c \approx -0.029424$. (a) Asset price $x(t)$ for $t \in [0,1400]$;
(b) bond price $z(t)$ for $t \in [0,1400]$; (c) a zoomed out bubble of
the asset price $x(t)$ for $t \in [495,510]$; (d) a zoomed out negative bubble
of the bond price $z(t)$ for $t \in [495,510]$.

\vskip 5mm

{\bf Figure 9}. Change in the behavior of the asset price $x(t)$, with the
initial conditions $x_0 = 3$, $z_0 = 0.1$, parameters $b = 1$, and varying $g$:
(a) $g = -0.176 > g_c$, where $g_c(b) = -0.176862964$. The asset price $x(t)$
oscillates without convergence as $t \ra \infty$; (b) a bubble of $x(t)$ for
$t \in [69,75]$, with the same parameters as in (a); (c) $g = -0.05$. The asset
price $x(t)$ displays periodic bubbles and crashes, as $t \ra \infty$.
(d) a bubble of the asset price $x(t)$ for $t \in [69,75]$, under the same
parameters as in (c).

\vskip 5mm

{\bf Figure 10}. Logarithmic behavior of the asset price for the parameters
$b = 1$, $g = -0.001$, and the initial conditions $x_0 = 3$ and $z_0 = 0.1$.
(a) The function $\ln x(t)$ oscillates without convergence for $t \ra \infty$;
(b) a bubble of $\ln x(t)$ for $t \in [66,84]$ under the same parameters as
in (a).

\vskip 5mm

{\bf Figure 11}. Logarithmic behavior of the asset price $x(t)$ for the
parameters $b = 0.4$, $g = -0.029$, and the initial conditions $x_0 = 1$ and
$z_0 = 0.1$. Here, $g_c(b) = -0.029424$. (a) The function $\ln x(t)$ oscillates
without convergence as $t \ra \infty$; (b) a bubble of $\ln x(t)$ for
$t \in [0,160]$, with the same parameters as in (a).

\vskip 5mm

{\bf Figure 12}. Expansion exponent $\Lbd(t)$, with the parameters $b = 0.4$,
$g = -0.029$, and the initial conditions $x_0 = 1$ and $z_0 = 0.1$, for different
temporal scales: (a) $t \in [100,1000]$; (b) $t \in [110,130]$. At
$t_{max}\approx 122.89$, the asset price $x(t)$ has its first local maximum
$x_{max} = 61.7171$, which corresponds to the first local minimum of $\Lbd(t)$.

\vskip 5mm

{\bf Figure 13}. Comparison of the numerical solution for the asset price
$x(t)$ (solid line) and its approximation $x_{app}(t)$ (dashed-dotted line),
with initial conditions $x_0 = 1$ and $z_0 = 0.1$, for different parameters
$b$ and $g$, where $1/e < b < b_0\approx 0.47$, and for different intervals 
of time: (a) $b = 0.4$, $g = -0.029 > g_c = -0.02943$, and $t \in [0,121]$;
(b) $b$ and $g$, as in Fig. 13a, but $t \in [110,121]$; (c) $b = 0.38$, 
$g = -0.0117 > g_c = -0.01174$, and $t \in [0,499.6]$; (d) $b$ and $g$, as 
in Fig. 13c, but $t \in [489,499.6]$;

\vskip 5mm

{\bf Figure 14}. Change in the behavior of the asset price $x(t)$, with fixed
$g = -0.03$, when varying $b$ in the vicinity of the boundary point
$b_2 = 0.400691$ separating the regions with different numbers of fixed
points. The initial conditions are $x_0 = 1$ and $z_0 = 0.1$.
(a) Asset price $x(t)$ for $b = 0.401 > b_2$ (solid line) and for $b = 0.41$
(dashed-dotted line) for $t \in [0,200]$; (b) asset price $x(t)$ for
$b = 0.401 > b_2$ (solid line) and for $b = 0.501$ (dashed-dotted line) for
$t \in [0,200]$; (c) asset price $x(t)$ for $b = 1$ (solid line), for $b = 2$
(dashed line), and for $b = 10$ (dashed-dotted line), with $t \in [0,20]$.

\vskip 5mm

{\bf Figure 15}. Asset price $x(t)$ for the parameter $g = -0.03$, but with
different $b$ and for different initial conditions $\{x_0,z_0\}$.
(a) Asset price $x(t)$ for $b = 0.4007 > b_2$ (solid line), $b = 0.4006 < b_2$
(dashed-dotted line), and the initial conditions $x_0 = 1$ and $z_0 = 0.1$.
Here, $b_2 = 0.400691$ is the boundary point, such that for $b > b_2$, there
is an unstable focus and a limit cycle, while for $b < b_2$, there are three
fixed points. The fixed point $\{x_3 = 2.928, z_1 = 0.916\}$ is a stable node,
with the Lyapunov exponents $\{\lbd_1 = -0.903,\lbd_2 = -0.022\}$.
The fixed point $\{x_2 = 3.08, z_2 = 0.912\}$ is a saddle, with the Lyapunov
exponents $\{\lbd_1 = -0.899,\lbd_2 = +0.022\}$, and the fixed point
$\{x_1 = 58.26, z_1 = 0.174\}$ is an unstable focus, with the Lyapunov
exponents $\{\lbd_1 = \lbd_2^* = +1.03 - i1.72\}$. (b) Asset price $x(t)$
for the parameters $g = -0.03$, $b = 0.4007 > b_2$, and different initial
conditions: $\{x_0 = 1, z_0 = 0.1\}$ (solid line), $\{x_0 = 5, z_0 = 0.5\}$
(dashed line), and $\{x_0 = 0.01, z_0 = 50\}$ (dashed-dotted line).

\vskip 5mm

{\bf Figure 16}. Critical behavior of bubble characteristics. The temporal
interval between two neighboring bubbles is $L$. The bubble amplitude is $A$.
The distance from the critical line is $\Dlt = b - b_2$ in (a) and $g$ in (b).
The initial conditions are taken as $x_0 = 1$ and $z_0 = 0.1$.
(a) Interval $L$ in logarithmic units, as a function of log-distance
$\ln |\Dlt| = \ln |b - b_2|$ to the critical line. Here, $g = - 0.03$ and
$b_2 = 0.400691$. (b) Bubble amplitude $A$ in logarithmic units, as a
function of $\ln|g|$ measuring the distance from the critical line $g = -0$,
with fixed $b = 1$. Here $g_c = -0.1769$, so that $\ln g_c = -1.7322$.

\newpage

\begin{center}
{\bf Table Caption}
\end{center}

{\bf Table 1}. Fast divergence of the bubble amplitude $A$, when decreasing $g$,
with fixed $b = 1$. $N$ is the number of bubbles in the time interval
$[0,100]$, and $w$ is the bubble width measured at half-amplitude.

\vskip 3cm

\begin{center}
{\bf Table 1}

\vskip 5mm

\begin{tabular}{|c|c|c|c|} \hline
$g$ & $N$ & $A$ & $w$ \\ \hline
$-0.1768$     & 19 & $0.115\cdot 10^2$ & 2.5 \\ \hline
$-0.1$        & 15 & $0.31 \cdot 10^2$ & 1.4 \\ \hline
$-10^{-2}$    & 8  & $0.63 \cdot 10^3$ & 0.9 \\ \hline
$-10^{-3}$    & 6  & $0.9  \cdot 10^4$ & 0.8 \\ \hline
$-10^{-4}$    & 5  & $1.2  \cdot 10^5$ & 0.75 \\ \hline
$-10^{-5}$    & 4  & $1.4  \cdot 10^6$ & 0.7 \\ \hline
$-10^{-6}$    & 3  & $1.7  \cdot 10^7$ & 0.7 \\ \hline
$-10^{-7}$    & 3  & $1.9  \cdot 10^8$ & 0.7 \\ \hline
$-10^{-8}$    & 2  & $2.1  \cdot 10^9$ & 0.7 \\ \hline
$-10^{-9}$    & 2  & $2.4 \cdot 10^{10}$ & 0.7 \\ \hline
$-10^{-10}$   & 2  & $2.6 \cdot 10^{11}$ & 0.7 \\ \hline
\end{tabular}
\end{center}

\newpage

\begin{figure}[ht]
\centerline{\includegraphics[width=12cm]{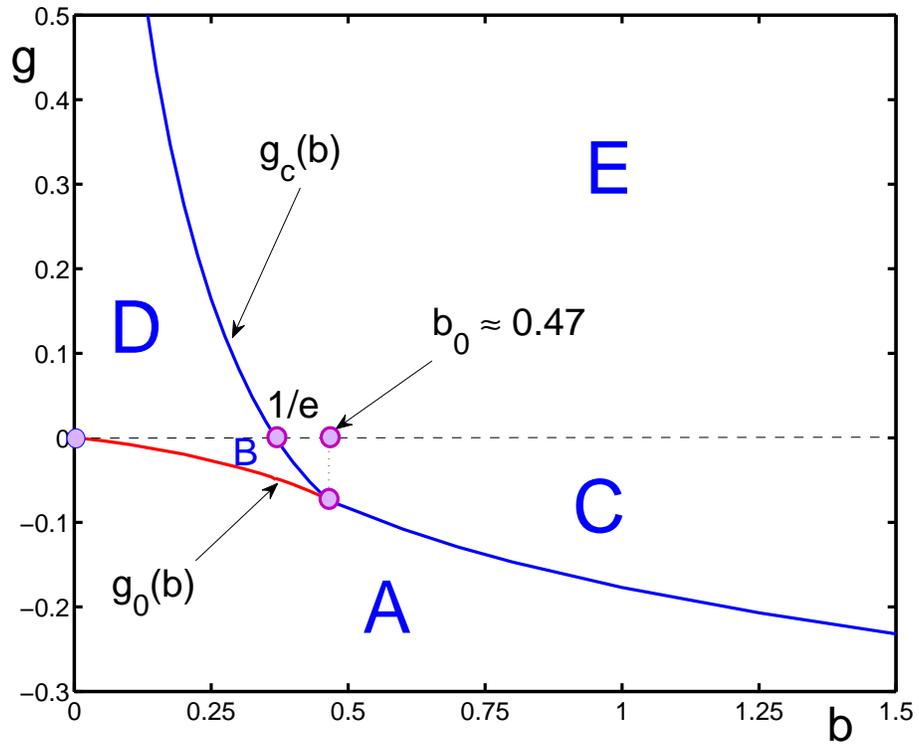} }
\caption{Region of existence of stationary solutions corresponding
to the fixed points of the evolution equations for the asset price $x$
and bond price $z$.
}
\label{fig:Fig.1}
\end{figure}

\newpage

\begin{figure}[ht]
\vspace{9pt}
\centerline{
\hbox{ \includegraphics[width=8.5cm]{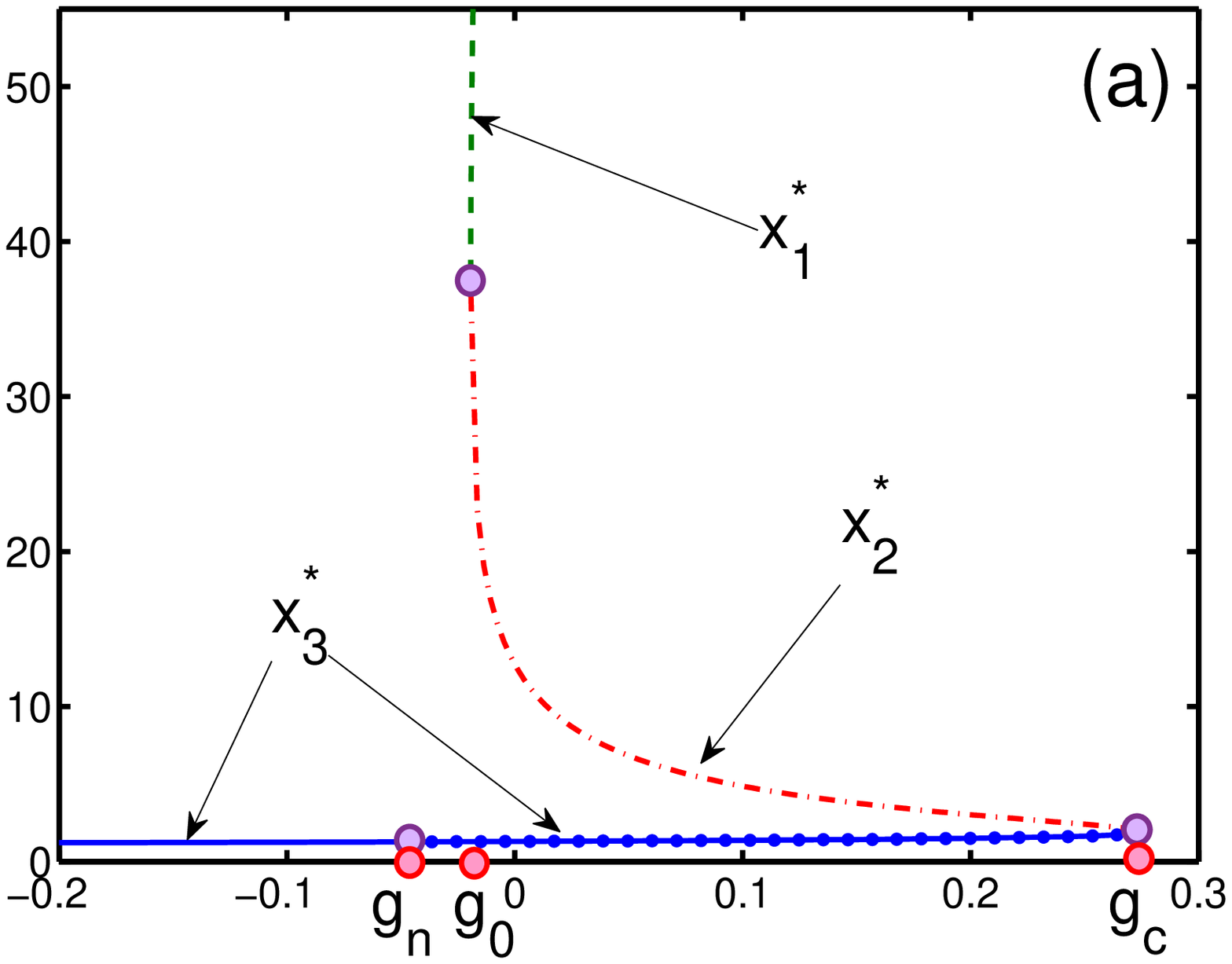} \hspace{2cm}
\includegraphics[width=8.5cm]{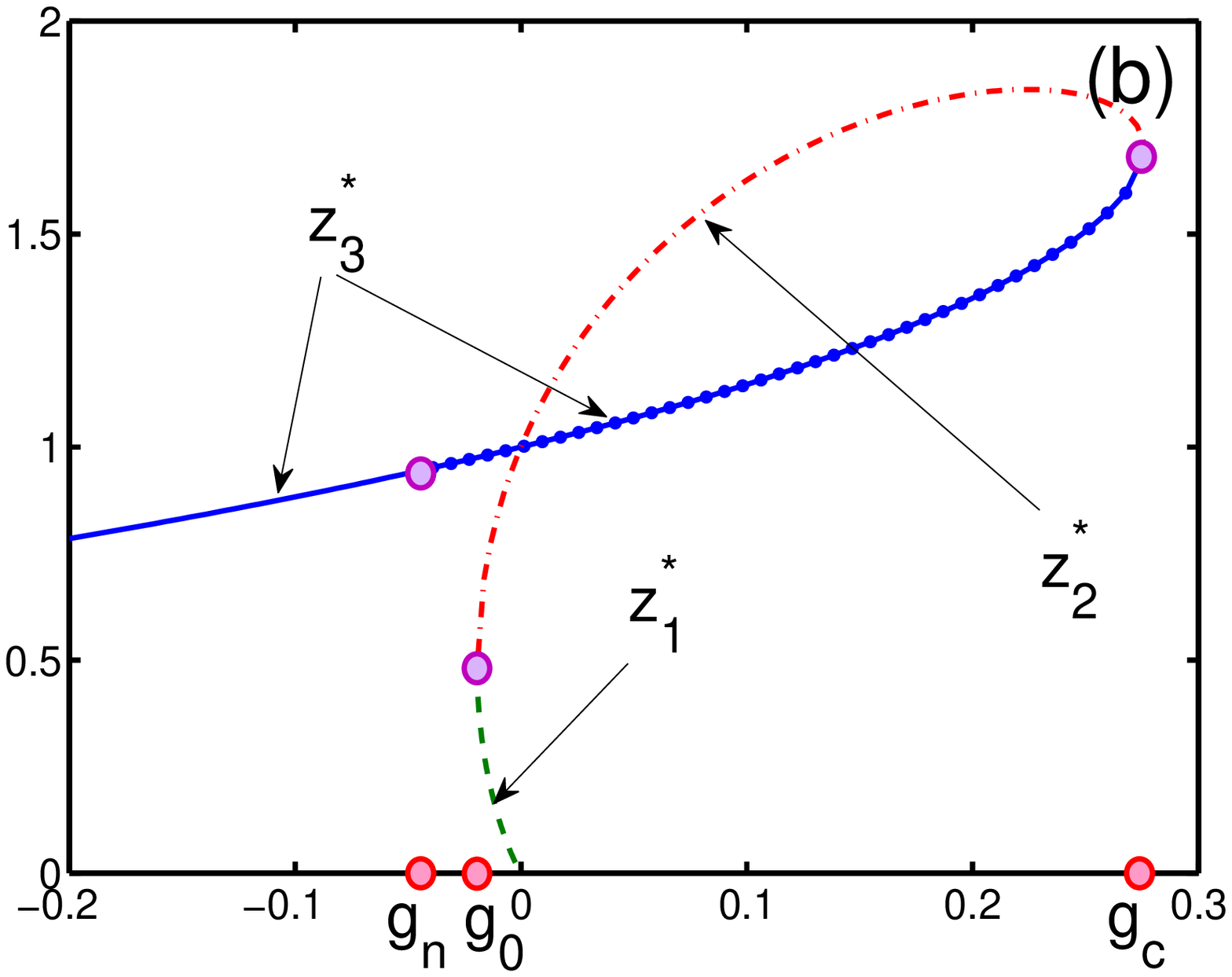} } }
\caption{Bifurcation path
${\bf A \longrightarrow B \longrightarrow D \longrightarrow E}$.
Here $b = 0.2 < 1/e$ is fixed and $g$ varies: (a) asset price $x^*$;
(b) bond price $z^*$. The fixed point $\{x_3^*,z_3^*\}$ (solid line)
is a stable focus transforming into a stable node (solid line with dots)
at $g_n \approx -0.0471$. The point exists for $- \infty < g < g_c$.
The fixed point $\{x_2^*,z_2^*\}$ (dashed-dotted line) is a saddle,
which exists for $g_0 < g < g_c$, and the fixed point $\{x_1^*,z_1^*\}$
(dashed line) is an unstable focus, which exists for $g_0 < g < 0$.
At $g = g_0$, the fixed points $\{x_1^*,z_1^*\}$ and $\{x_2^*,z_2^*\}$
coincide. At $g = g_c$, the fixed points $\{x_3^*,z_3^*\}$ and
$\{x_2^*,z_2^*\}$ coincide. When $g \ra -0$, then $x_1^* \ra +\infty$
and $z_1^* \ra 0$. When $g \ra -\infty$, then $x_3^*\ra 1$ and $z_3^*\ra 0$.
When $g > g_c$, fixed points do not exist.
}
\label{fig:Fig.2}
\end{figure}

\newpage

\begin{figure}[ht]
\vspace{9pt}
\centerline{
\hbox{ \includegraphics[width=8.5cm]{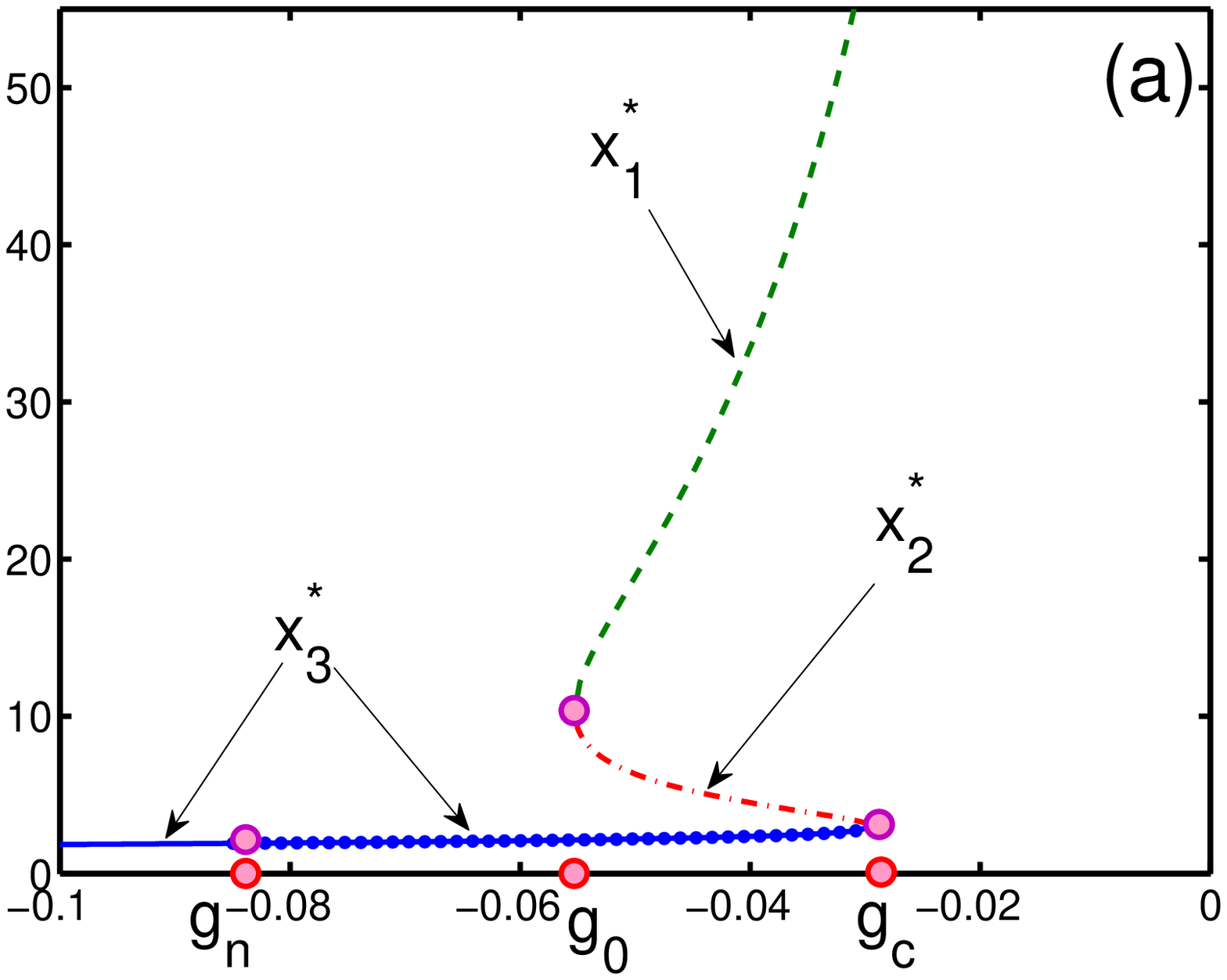} \hspace{2cm}
\includegraphics[width=8.5cm]{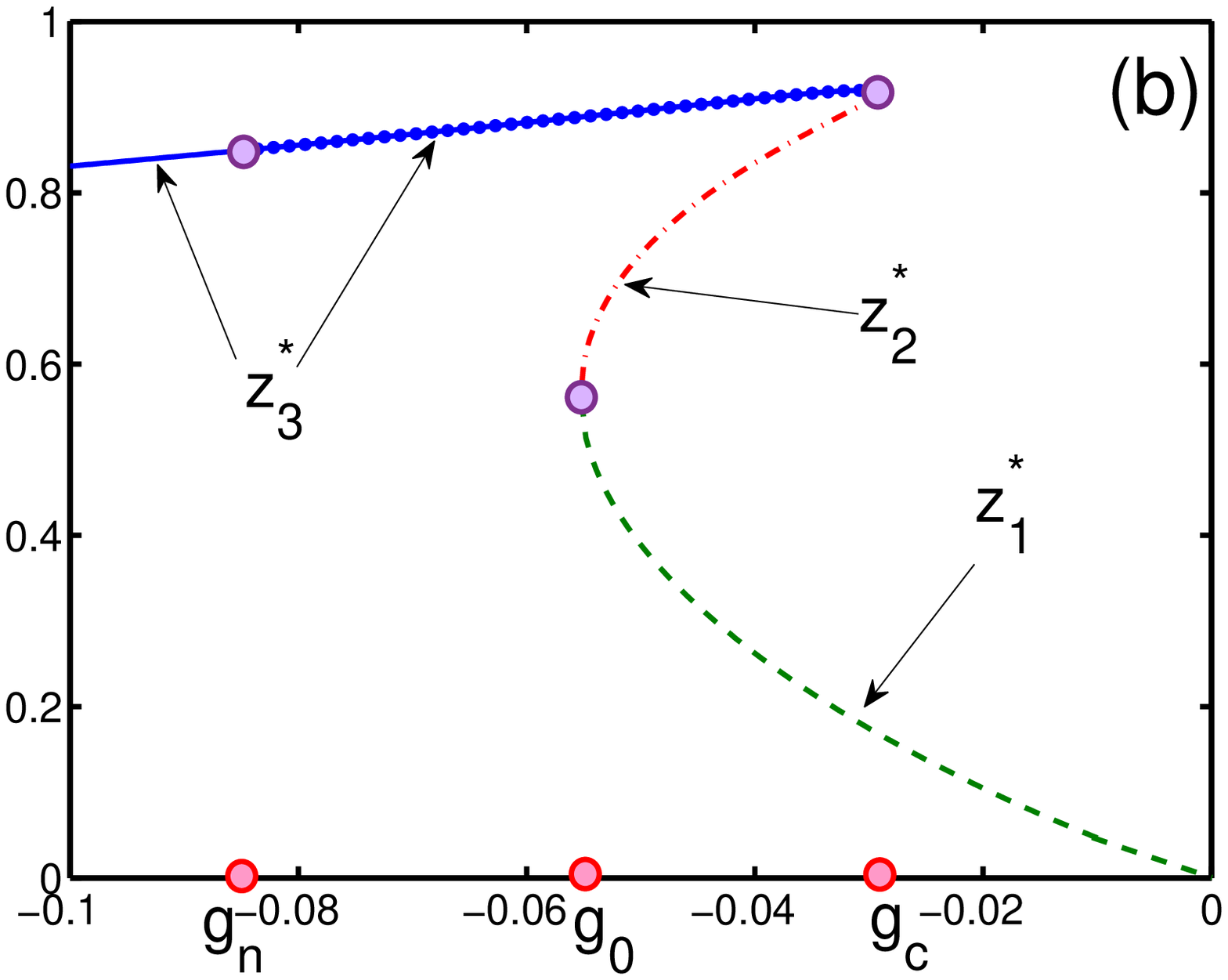} } }
\caption{Bifurcation path
${\bf A \longrightarrow B \longrightarrow C \longrightarrow E}$.
Here $1/e < b = 0.4 < b_0$ and $g$ varies: (a) asset price $x^*$;
(b) bond price $z^*$. The fixed point $\{x_3^*,z_3^*\}$ (solid line)
is a stable focus transforming to a stable node (solid line with dots) at
$g_n \approx -0.0849$. The point exists for $- \infty < g < g_c < 0$.
The fixed point $\{x_2^*,z_2^*\}$ (dashed-dotted line) is a saddle,
which exists for $g_0 < g < g_c$, and the fixed point $\{x_1^*,z_1^*\}$
(dashed line) is an unstable focus, which exists for $g_0 < g < 0$.
At $g = g_0$, the fixed points $\{x_1^*,z_1^*\}$ and $\{x_2^*,z_2^*\}$
coincide. At $g = g_c$, the fixed points $\{x_3^*,z_3^*\}$ and
$\{x_2^*,z_2^*\}$ coincide. When $g \ra -0$, then $x_1^* \ra +\infty$
and $z_1^* \ra 0$. When $g \ra - \infty$, then $x_3^* \ra 1$ and
$z_3^* \ra 0$. When $g \geq 0$, fixed points do not exist.
}
\label{fig:Fig.3}
\end{figure}

\newpage

\begin{figure}[ht]
\vspace{9pt}
\centerline{
\hbox{ \includegraphics[width=8.5cm]{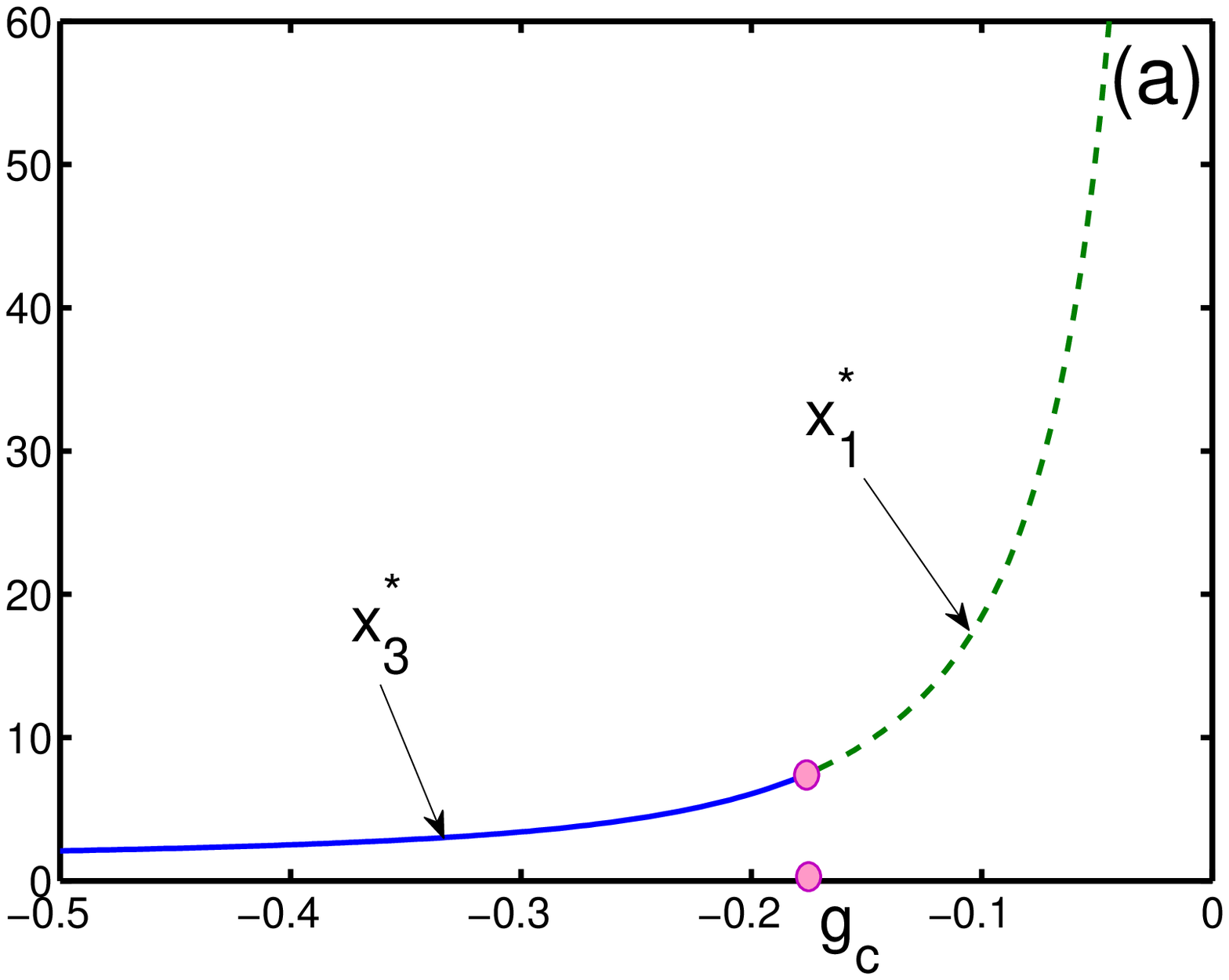} \hspace{2cm}
\includegraphics[width=8.5cm]{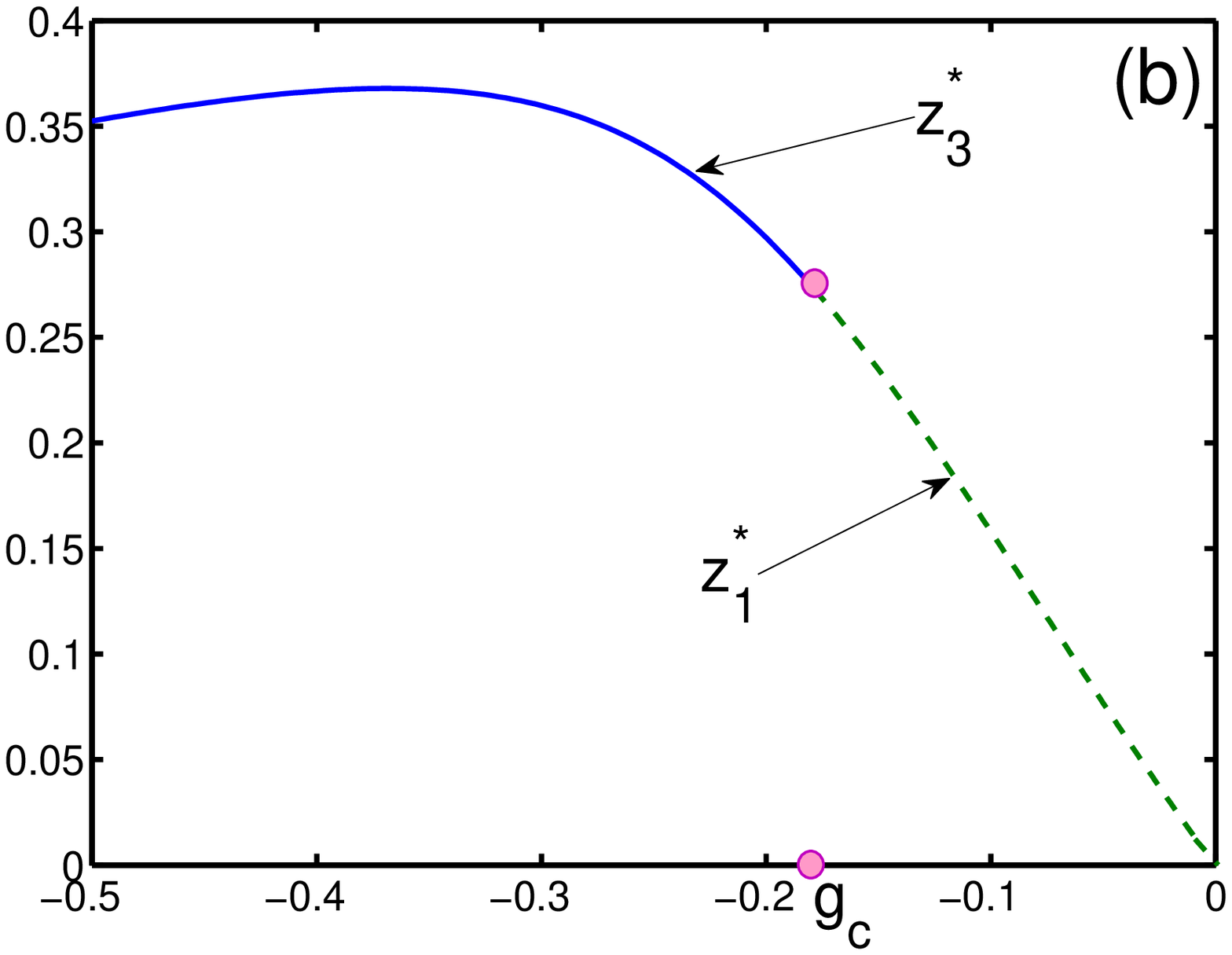} } }
\caption{Bifurcation path ${\bf A \longrightarrow C \longrightarrow E}$.
Here $b = 1 > b_0$ is fixed and $g$ varies: (a) asset price $x^*$;
(b) bond price $z^*$. The fixed point $\{x_3^*,z_3^*\}$ (solid line)
is a stable focus, and the fixed point $\{x_1^*,z_1^*\}$ (dashed line) is
an unstable focus. At $g = g_c \approx -0.1769$, the stable focus transforms
into an unstable focus with $x_3^* = x_1^* = e^2$, $z_3^* = z_1^* = 2/(be^2)$,
the Lyapunov exponents being ${\Re} \lbd_{1,2} = 0$.
}
\label{fig:Fig.4}
\end{figure}

\newpage

\begin{figure}[ht]
\vspace{9pt}
\centerline{
\hbox{ \includegraphics[width=8.5cm]{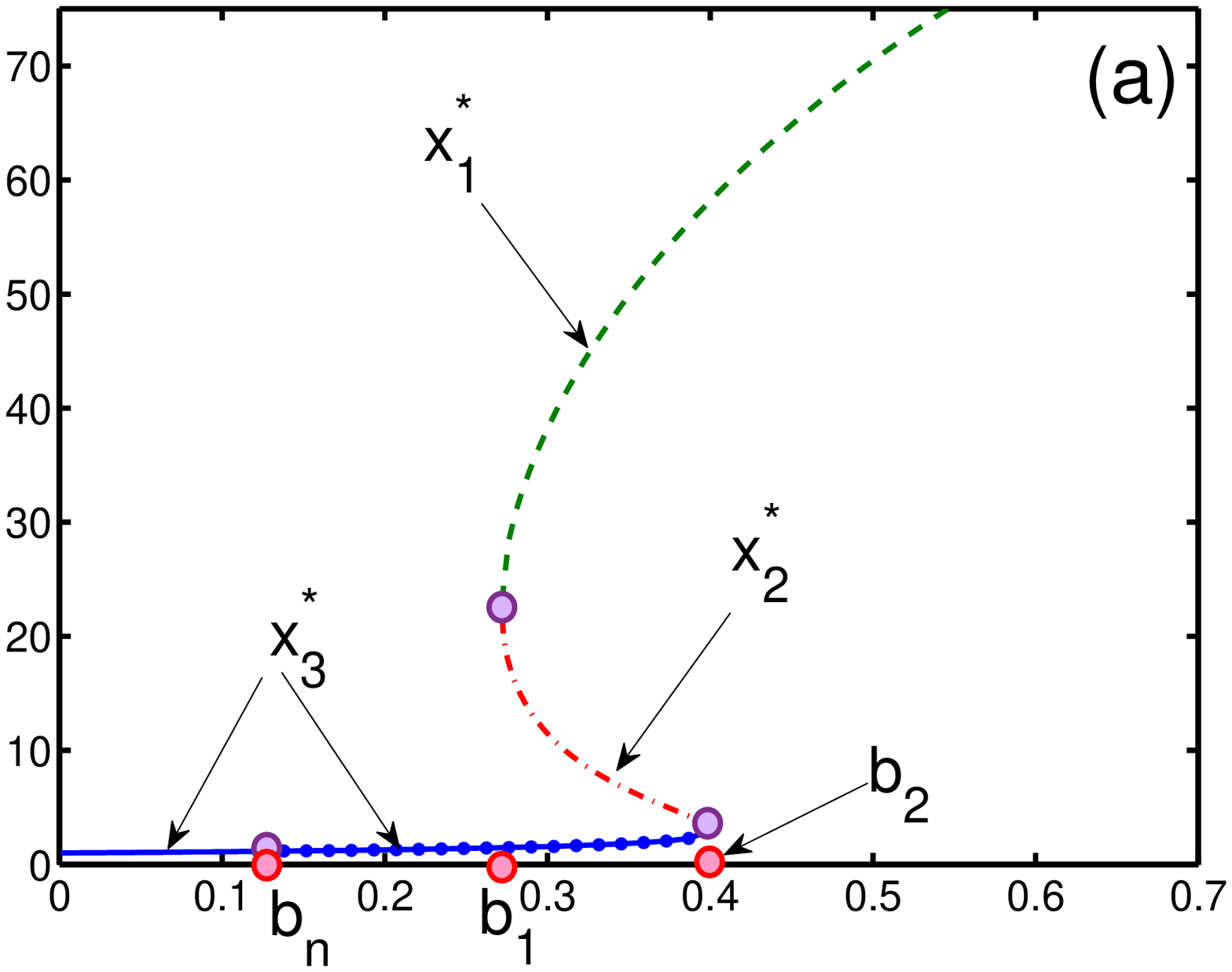} \hspace{2cm}
\includegraphics[width=8.5cm]{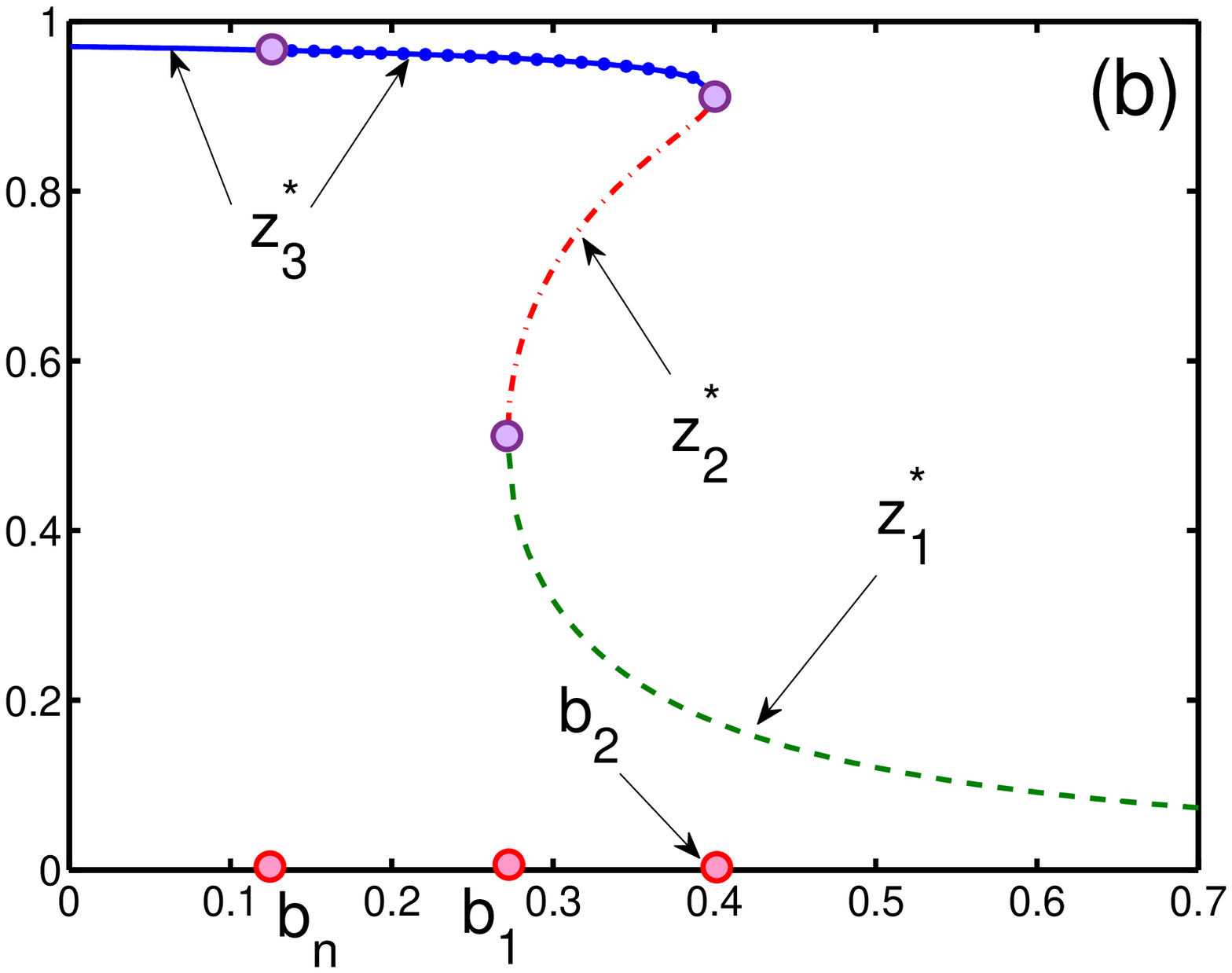} } }
\caption{Bifurcation path ${\bf A \longrightarrow B \longrightarrow C}$.
Here $g = -0.03$ is fixed and $b$ varies: (a) asset price $x^*$;
(b) bond price $z^*$. The fixed point $\{x_3^*,z_3^*\}$ (solid line)
is a stable focus transforming to a stable node (solid line with dots) at
$b_n \approx 0.1242$. The point exists for $0 \leq b < b_2$, where
$b_2 \approx 0.4007$. The fixed point $\{x_2^*,z_2^*\}$ (dashed-dotted line)
is a saddle, which exists for $b_1 < b < b_2$, where $b_1 \approx 0.2718$.
The fixed point $\{x_1^*,z_1^*\}$ (dashed line) is an unstable focus, which
exists for $b > b_1$. At $b = b_1$, the fixed points $\{x_1^*,z_1^*\}$ and
$\{x_2^*,z_2^*\}$ coincide. At $b = b_2$, the fixed points $\{x_3^*,z_3^*\}$
and $\{x_2^*,z_2^*\}$ coincide. When $b = 0$, then $x_3^* = 1$ and $z_3^* = e^g$.
When $b \ra +\infty$, then $x_1^* \ra +\infty$ and $z_1^* \ra 0$.
}
\label{fig:Fig.5}
\end{figure}

\newpage

\begin{figure}[ht]
\vspace{9pt}
\centerline{
\hbox{ \includegraphics[width=8.5cm]{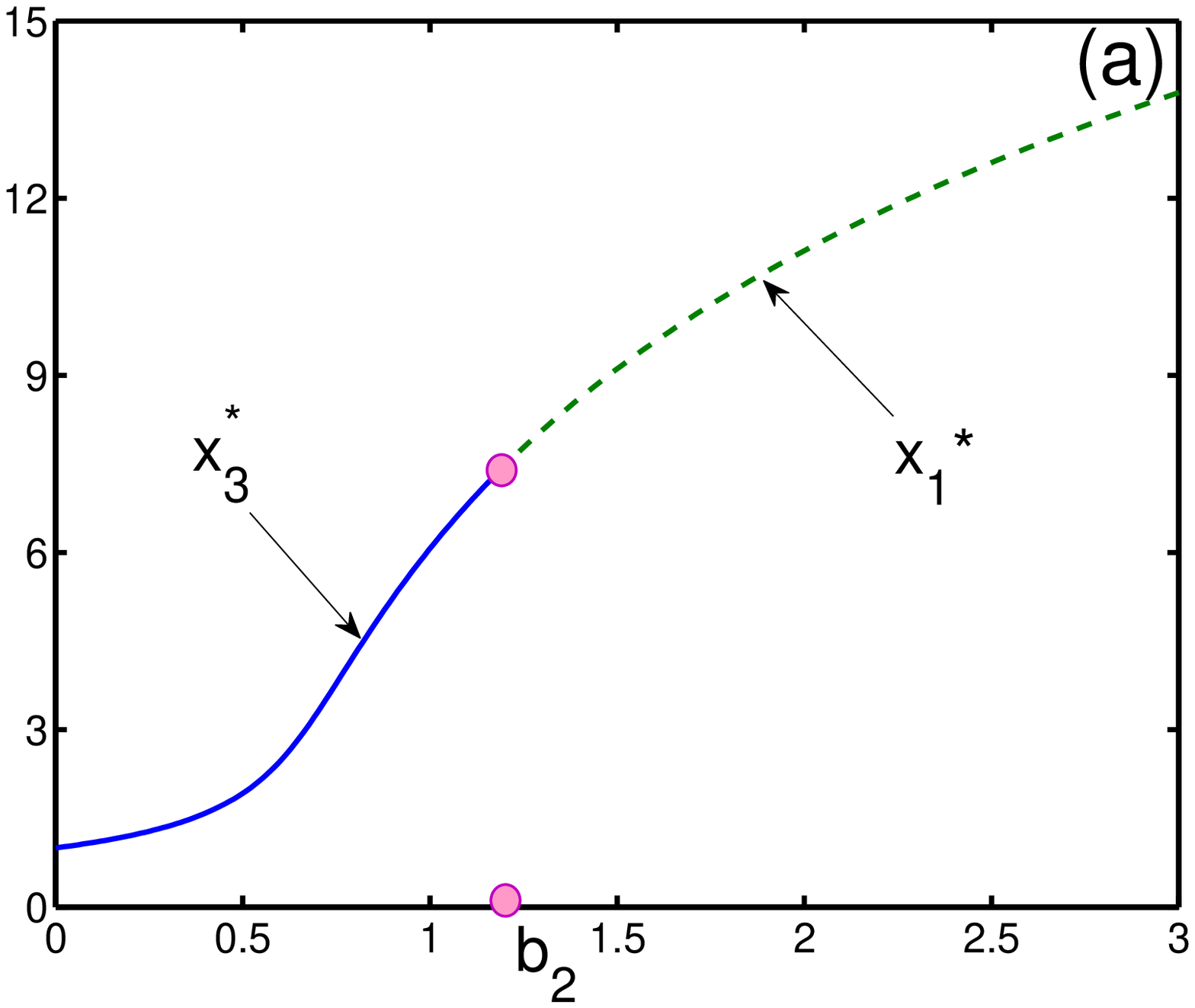} \hspace{1cm}
\includegraphics[width=8.5cm]{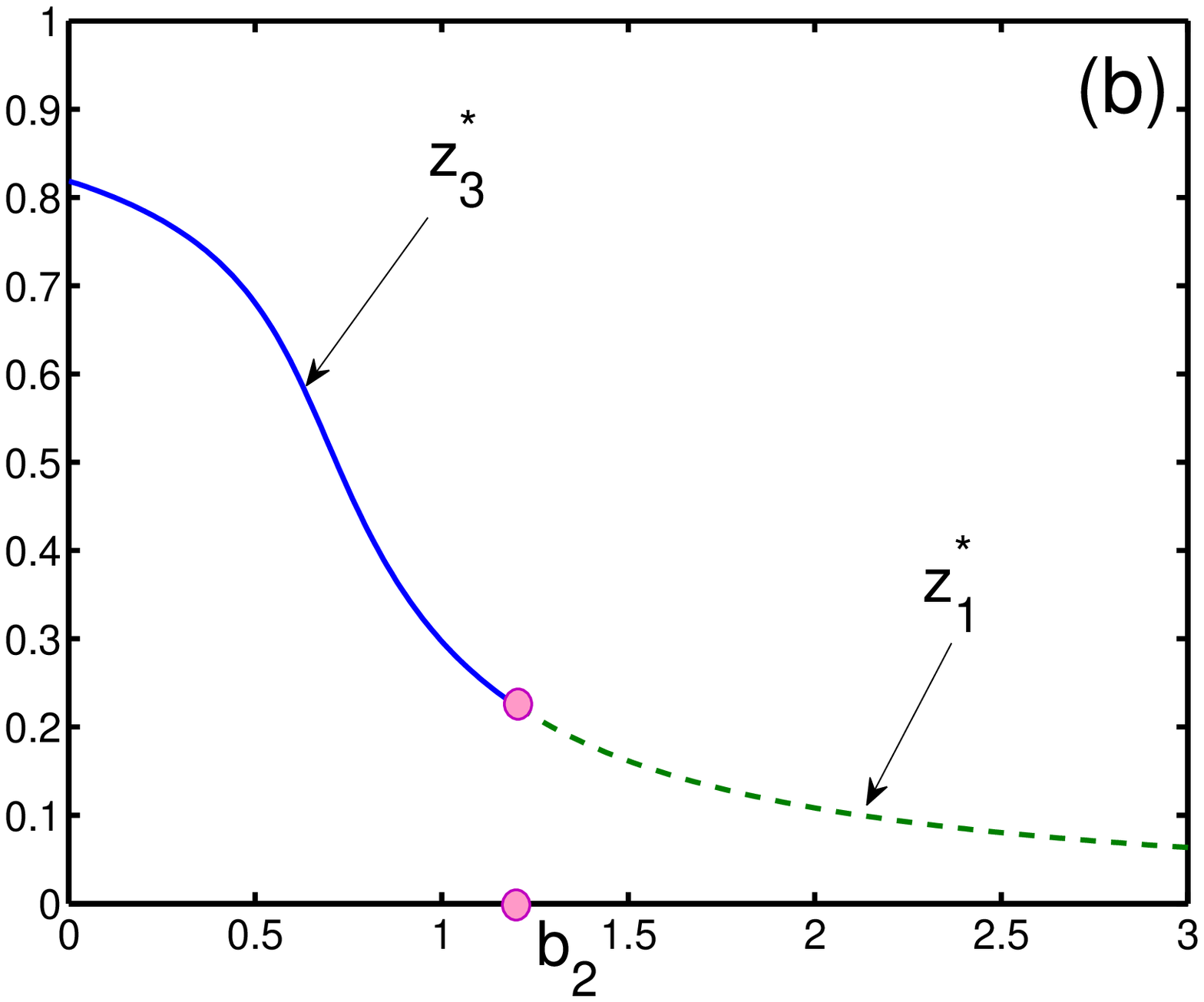} } }
\caption{Bifurcation path ${\bf A \longrightarrow C}$. Variation of fixed
points under fixed $g = -0.2$ and changing $b$: (a) asset price $x^*$;
(b) bond price $z^*$. The fixed point $\{x_3^*,z_3^*\}$ (solid line)
is a stable focus that at $b = b_2 \approx 1.1864$ transforms into an unstable
focus $\{x_1^*,z_1^*\}$ (dashed line). At $b = b_2$, the fixed points
$\{x_3^*,z_3^*\}$ and $\{x_1^*,z_1^*\}$ coincide, with the Lyapunov exponents
being ${\Re}\lbd_{1,2} = 0$.
}
\label{fig:Fig.6}
\end{figure}

\newpage

\begin{figure}[ht]
\vspace{9pt}
\centerline{
\hbox{ \includegraphics[width=8.5cm]{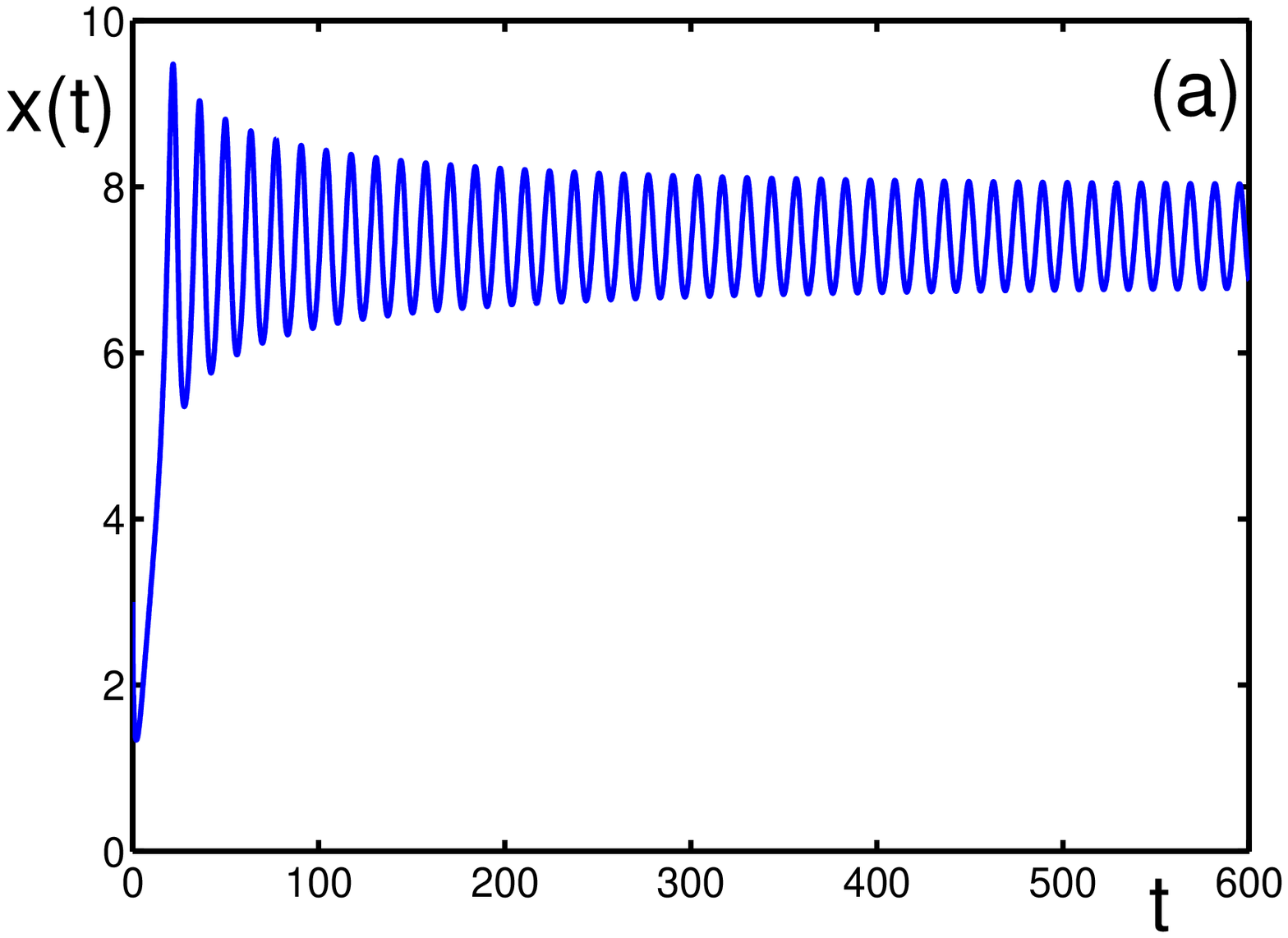} \hspace{1cm}
\includegraphics[width=8.5cm]{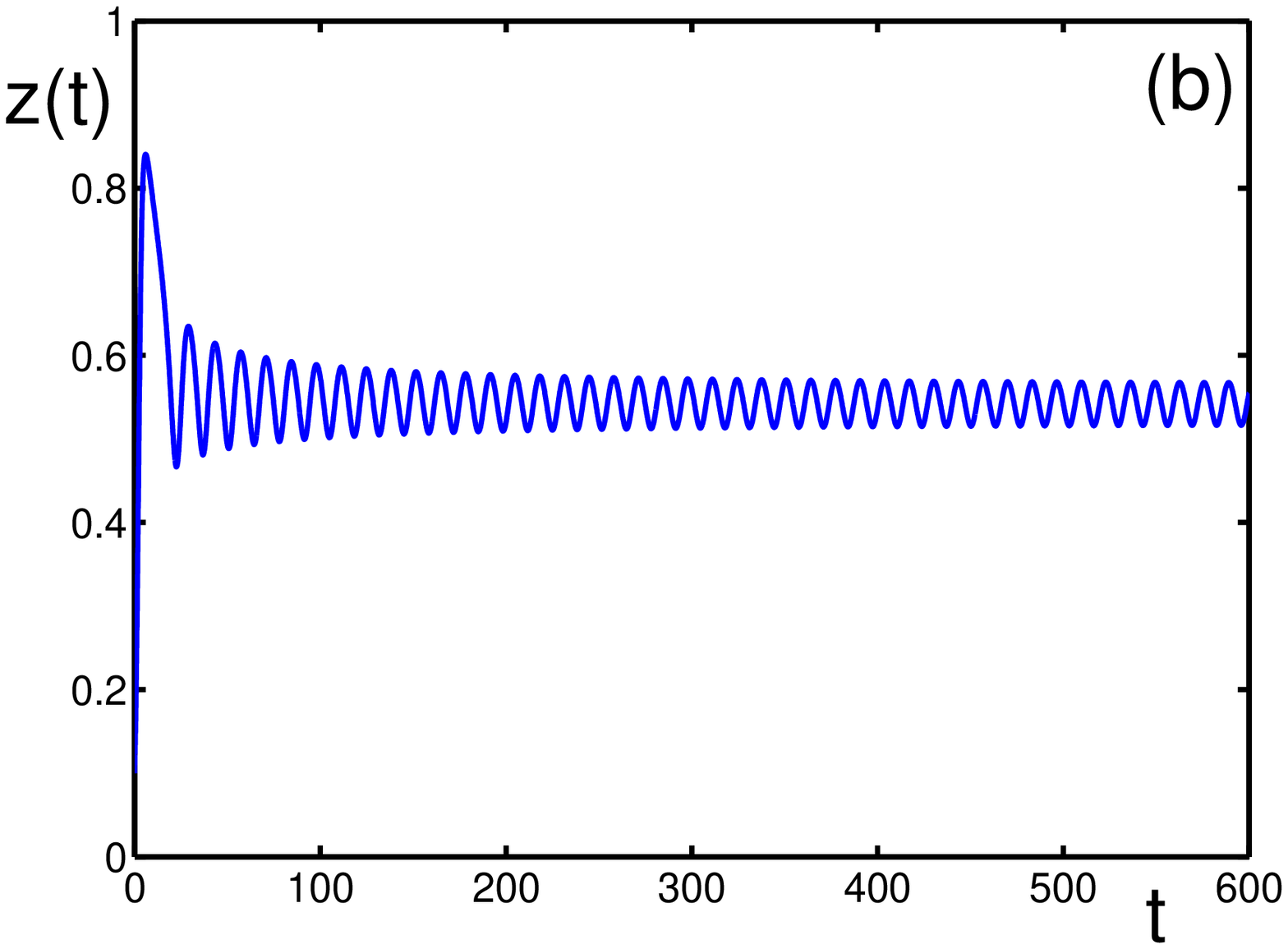}  } }
\vspace{9pt}
\centerline{
\hbox{ \includegraphics[width=8.5cm]{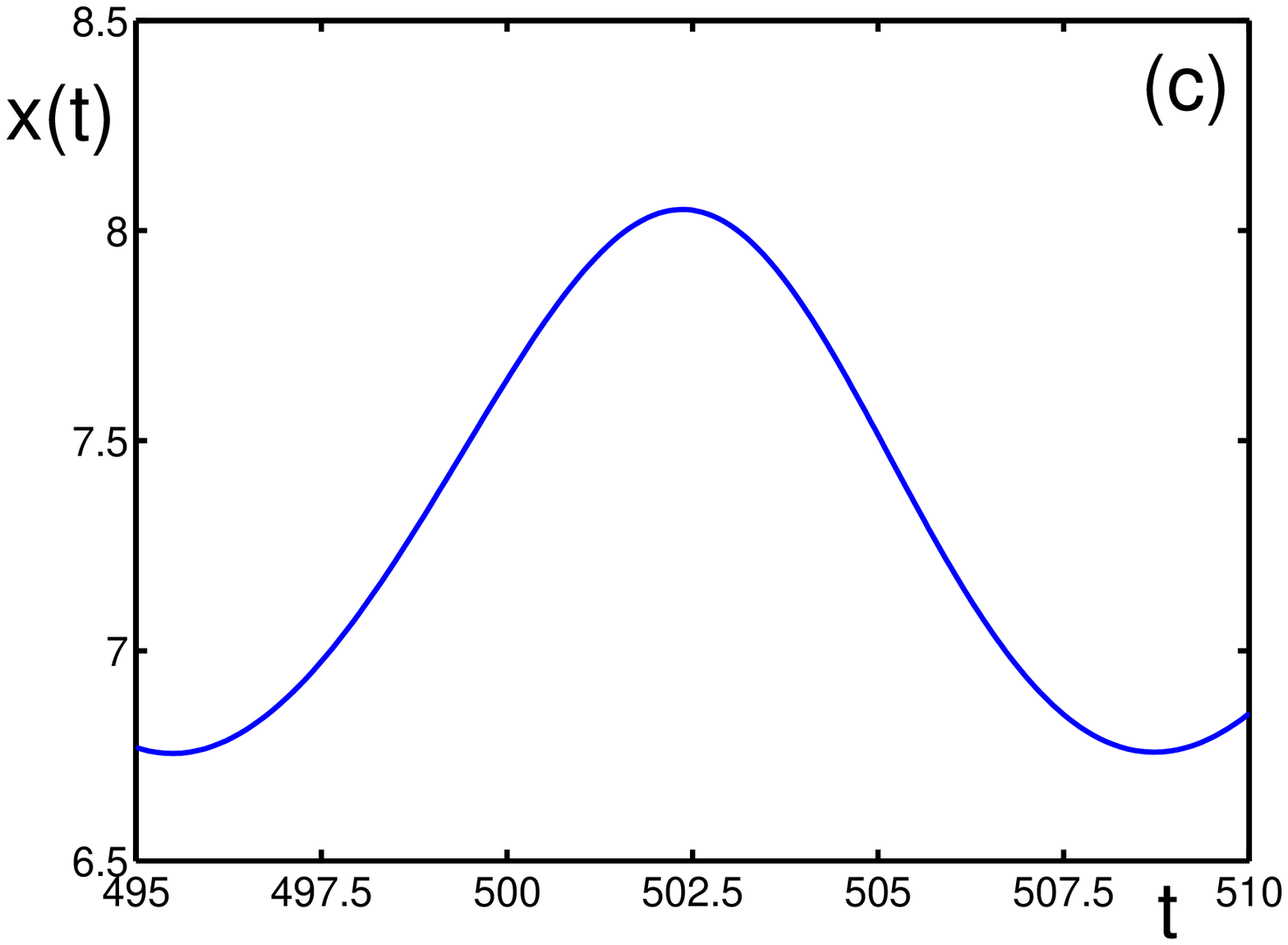} \hspace{1cm}
\includegraphics[width=8.5cm]{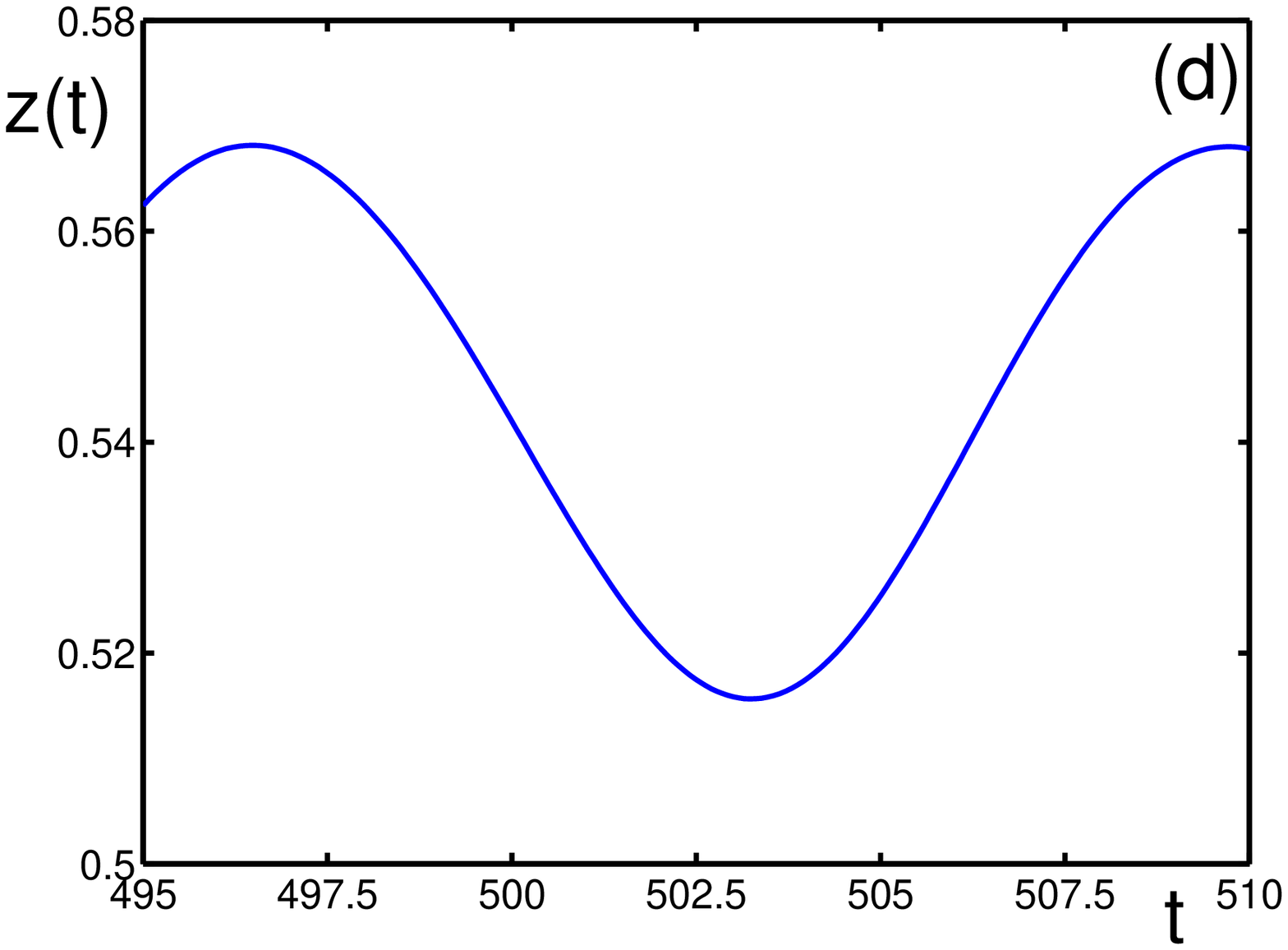} } }
\caption{Behavior of the solutions for the asset price and bond price
for the parameters $b = 0.5$, $g = -0.083$, and the initial conditions $x_0 = 3$
and $z_0 = 0.1$. Here, $g_c \approx -0.083056$. (a) Asset price $x(t)$ for
$t \in [0,600]$; (b) bond price $z(t)$ for $t \in [0,600]$; (c) a single
zoomed out bubble of the asset price $x(t)$ for $t \in [495,510]$; (d) a zoomed
out negative bubble of the bond price $z(t)$ for $t \in [495,510]$.
}
\label{fig:Fig.7}
\end{figure}

\newpage

\begin{figure}[ht]
\vspace{9pt}
\centerline{
\hbox{ \includegraphics[width=8.5cm]{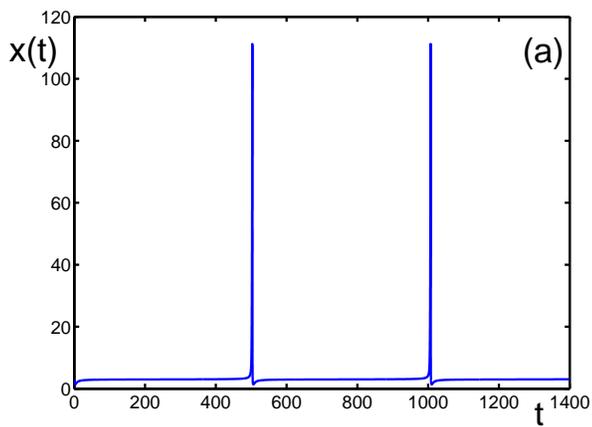} \hspace{1cm}
\includegraphics[width=8.5cm]{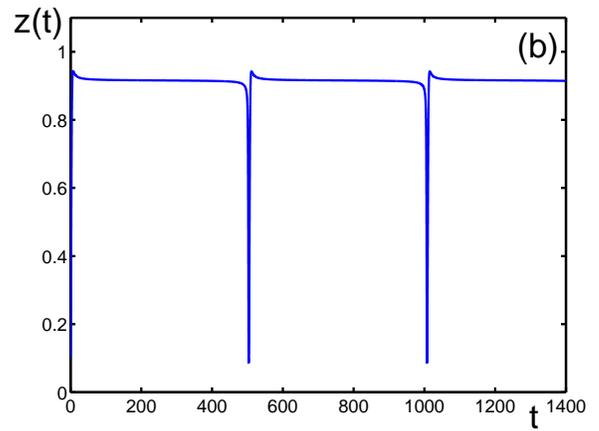}  } }
\vspace{9pt}
\centerline{
\hbox{ \includegraphics[width=8.5cm]{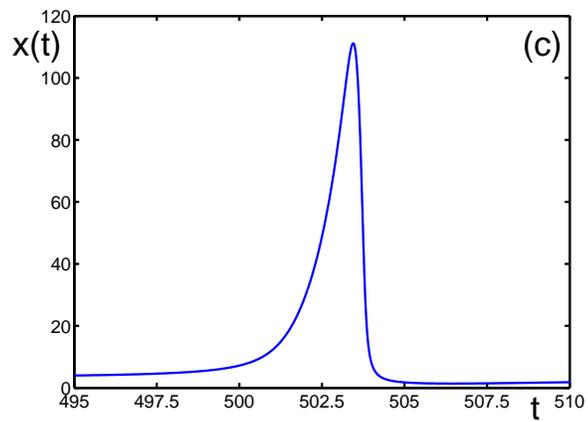} \hspace{1cm}
\includegraphics[width=8.5cm]{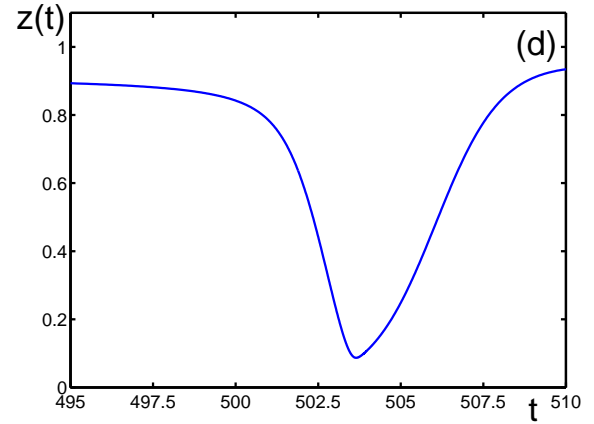} } }
\caption{Asset price and bond price for the parameters $b = 0.4$,
$g = -0.0294$, and the initial conditions $x_0 = 3$ and $z_0 = 0.1$. Here,
$g_c \approx -0.029424$. (a) Asset price $x(t)$ for $t \in [0,1400]$;
(b) bond price $z(t)$ for $t \in [0,1400]$; (c) a zoomed out bubble of
the asset price $x(t)$ for $t \in [495,510]$; (d) a zoomed out negative bubble of
the bond price $z(t)$ for $t \in [495,510]$.
}
\label{fig:Fig.8}
\end{figure}

\newpage

\begin{figure}[ht]
\vspace{9pt}
\centerline{
\hbox{ \includegraphics[width=8.5cm]{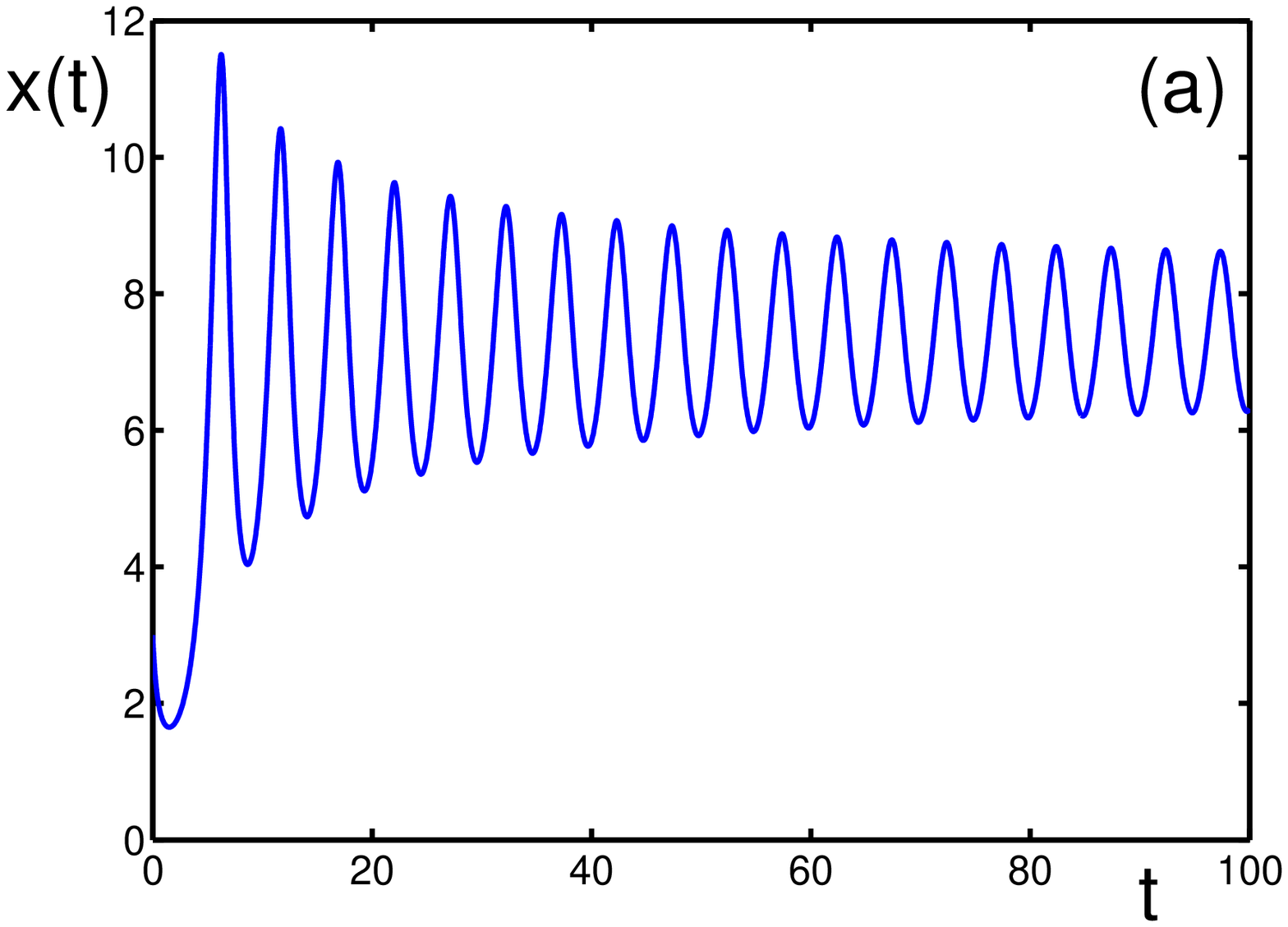} \hspace{1cm}
\includegraphics[width=8.5cm]{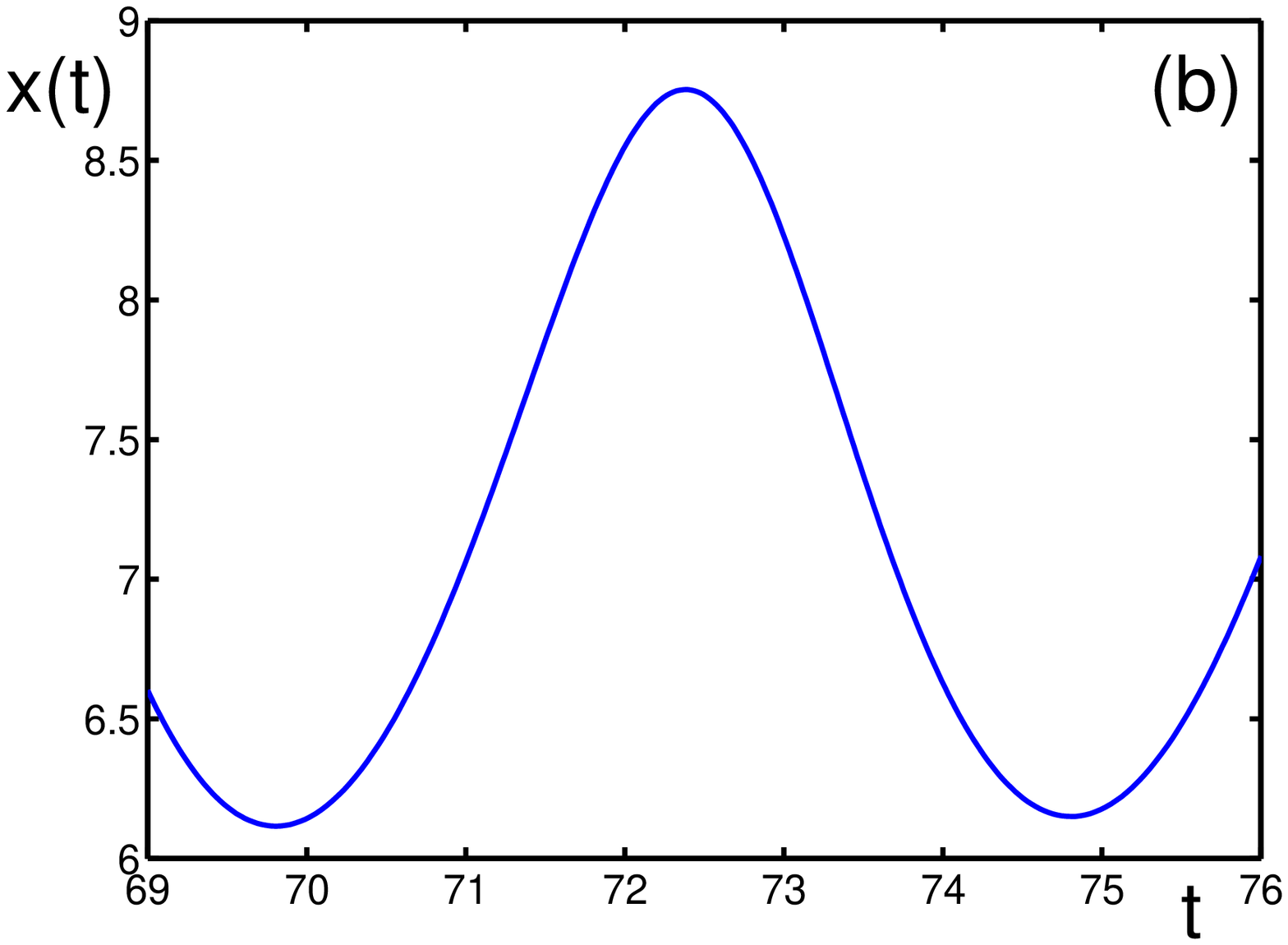}  } }
\vspace{9pt}
\centerline{
\hbox{ \includegraphics[width=8.5cm]{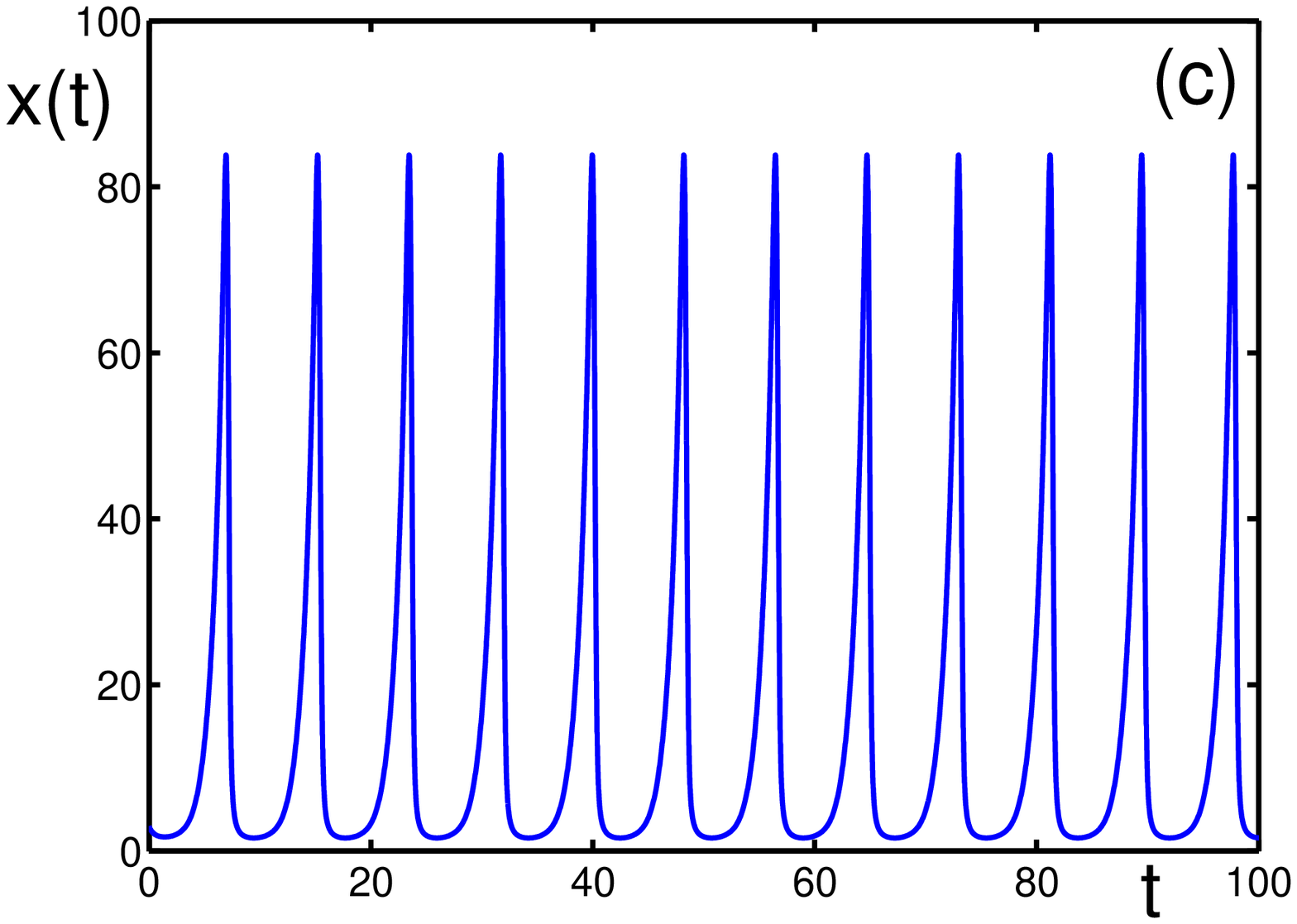} \hspace{1cm}
\includegraphics[width=8.5cm]{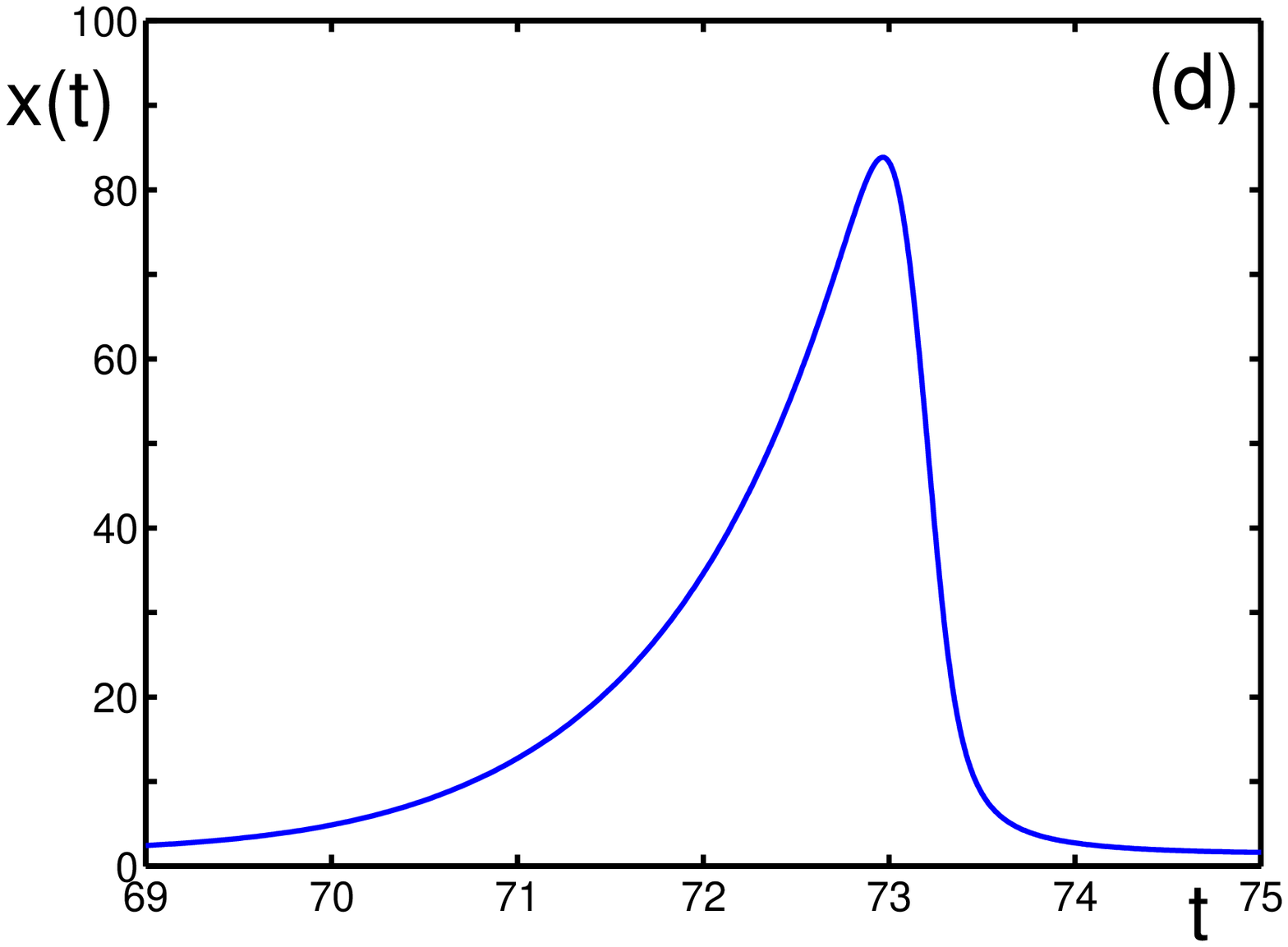} } }
\caption{Change in the behavior of the asset price $x(t)$, with the initial
conditions $x_0 = 3$, $z_0 = 0.1$, parameters $b = 1$, and varying $g$:
(a) $g = -0.176 > g_c$, where $g_c(b) = -0.176862964$. The asset price $x(t)$
oscillates without convergence as $t \ra \infty$; (b) a bubble of $x(t)$ for
$t \in [69,75]$, with the same parameters as in (a); (c) $g = -0.05$. The asset
price $x(t)$ displays periodic bubbles and crashes, as $t \ra \infty$.
(d) a bubble of the asset price $x(t)$ for $t \in [69,75]$, under the same
parameters as in (c).
}
\label{fig:Fig.9}
\end{figure}

\newpage

\begin{figure}[ht]
\vspace{9pt}
\centerline{
\hbox{ \includegraphics[width=8.5cm]{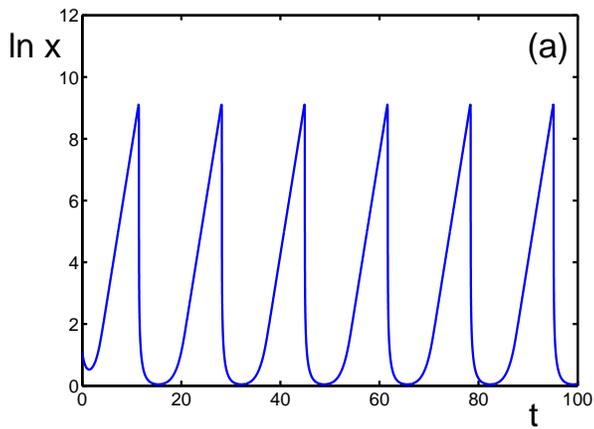} \hspace{1cm}
\includegraphics[width=8.5cm]{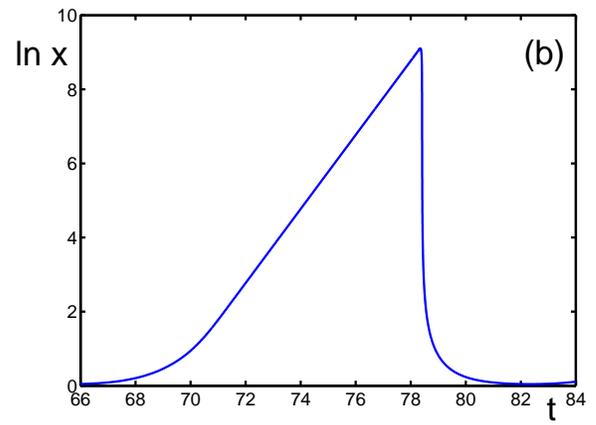} } }
\caption{Logarithmic behavior of the asset price for the parameters $b = 1$,
$g = -0.001$, and the initial conditions $x_0 = 3$ and $z_0 = 0.1$.
(a) The function $\ln x(t)$ oscillates without convergence for $t \ra \infty$;
(b) a bubble of $\ln x(t)$ for $t \in [66,84]$ under the same parameters as
in (a).
}
\label{fig:Fig.10}
\end{figure}

\newpage

\begin{figure}[ht]
\vspace{9pt}
\centerline{
\hbox{ \includegraphics[width=8.5cm]{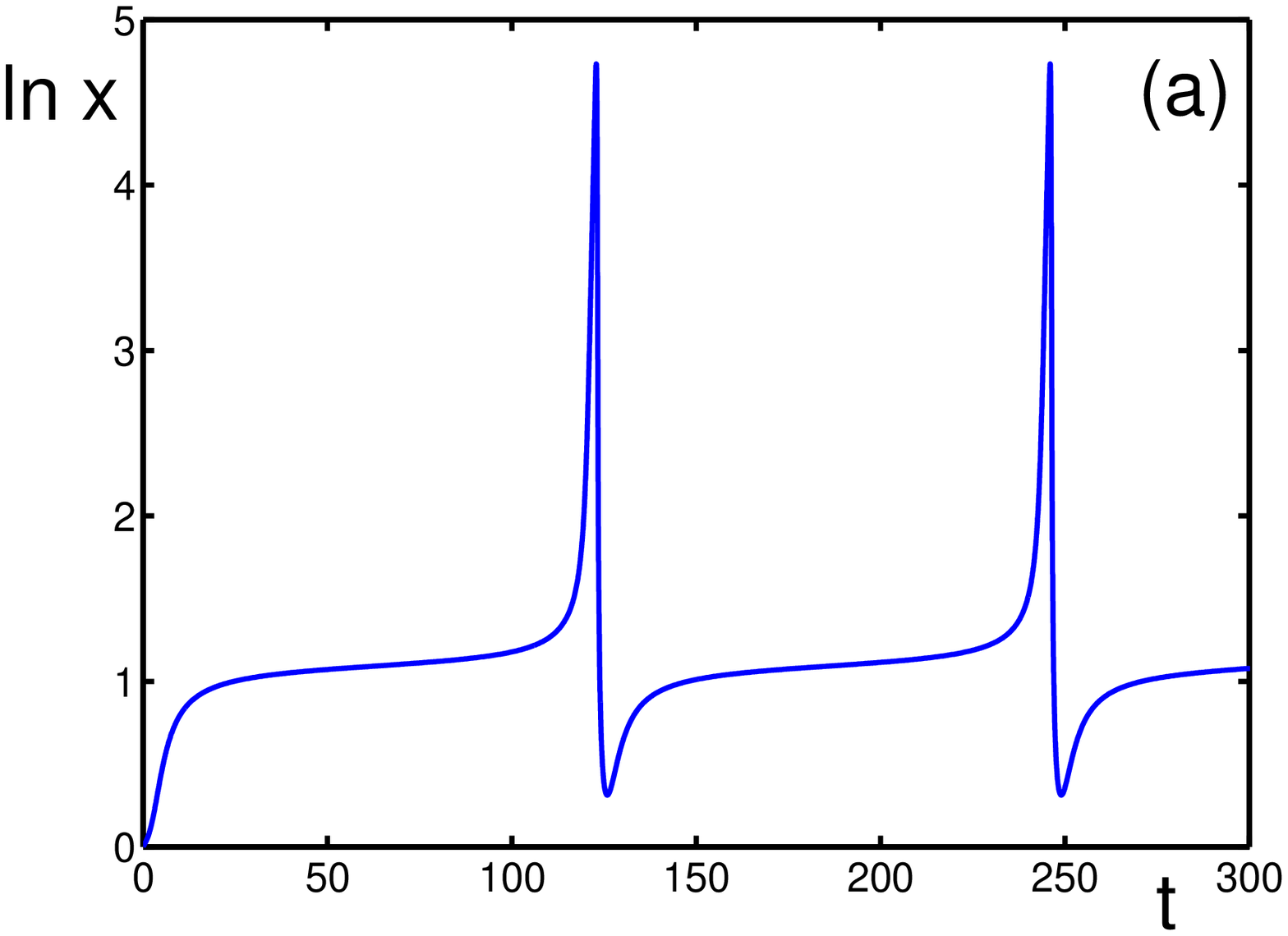} \hspace{1cm}
\includegraphics[width=8.5cm]{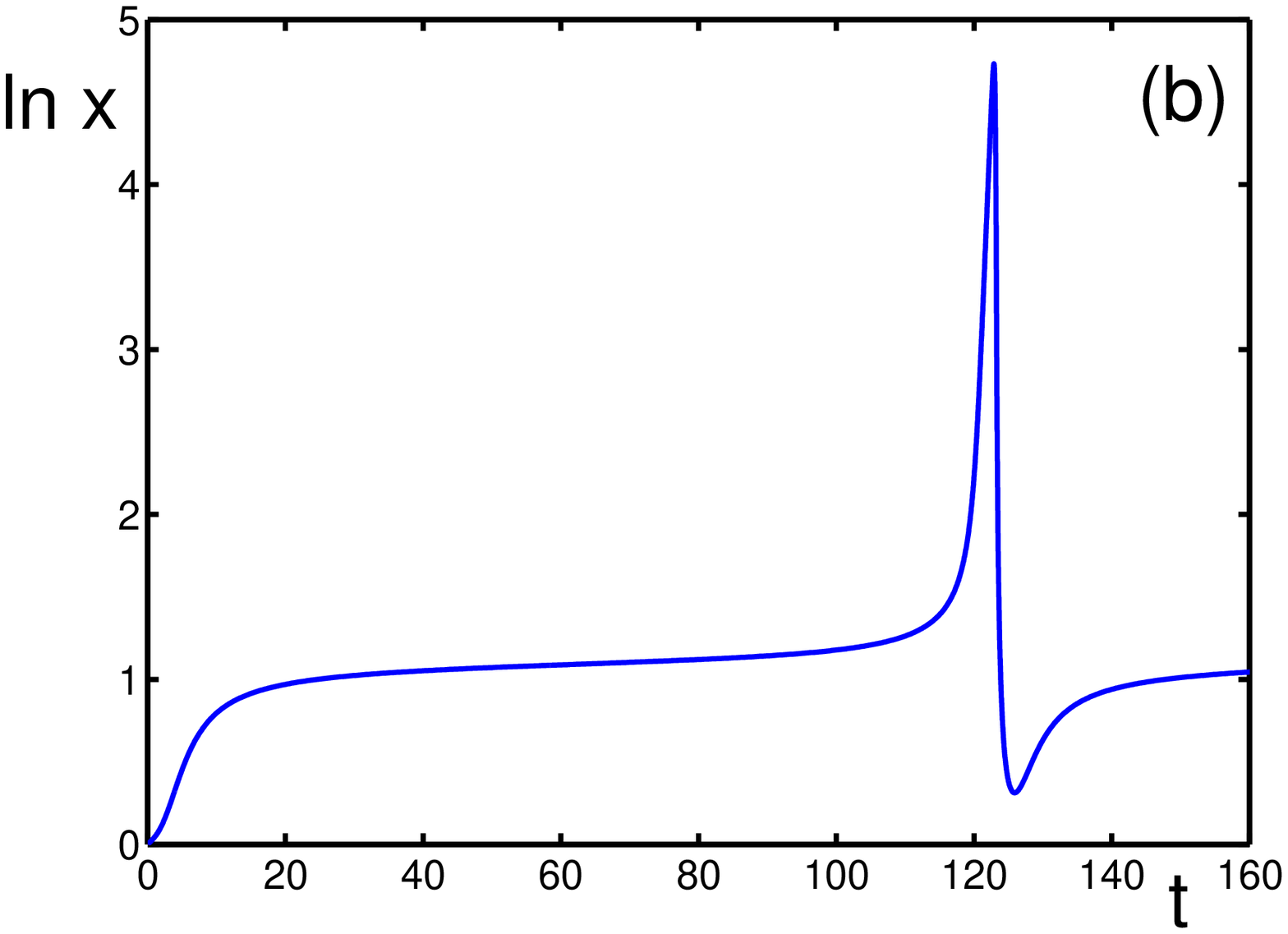} } }
\caption{Logarithmic behavior of the asset price $x(t)$ for the parameters $b = 0.4$,
$g = -0.029$, and the initial conditions $x_0 = 1$ and $z_0 = 0.1$. Here,
$g_c(b) = -0.029424$. (a) The function $\ln x(t)$ oscillates without convergence
as $t \ra \infty$; (b) a bubble of $\ln x(t)$ for $t \in [0,160]$, with the same
parameters as in (a).
}
\label{fig:Fig.11}
\end{figure}

\newpage

\begin{figure}[ht]
\vspace{9pt}
\centerline{
\hbox{ \includegraphics[width=8.5cm]{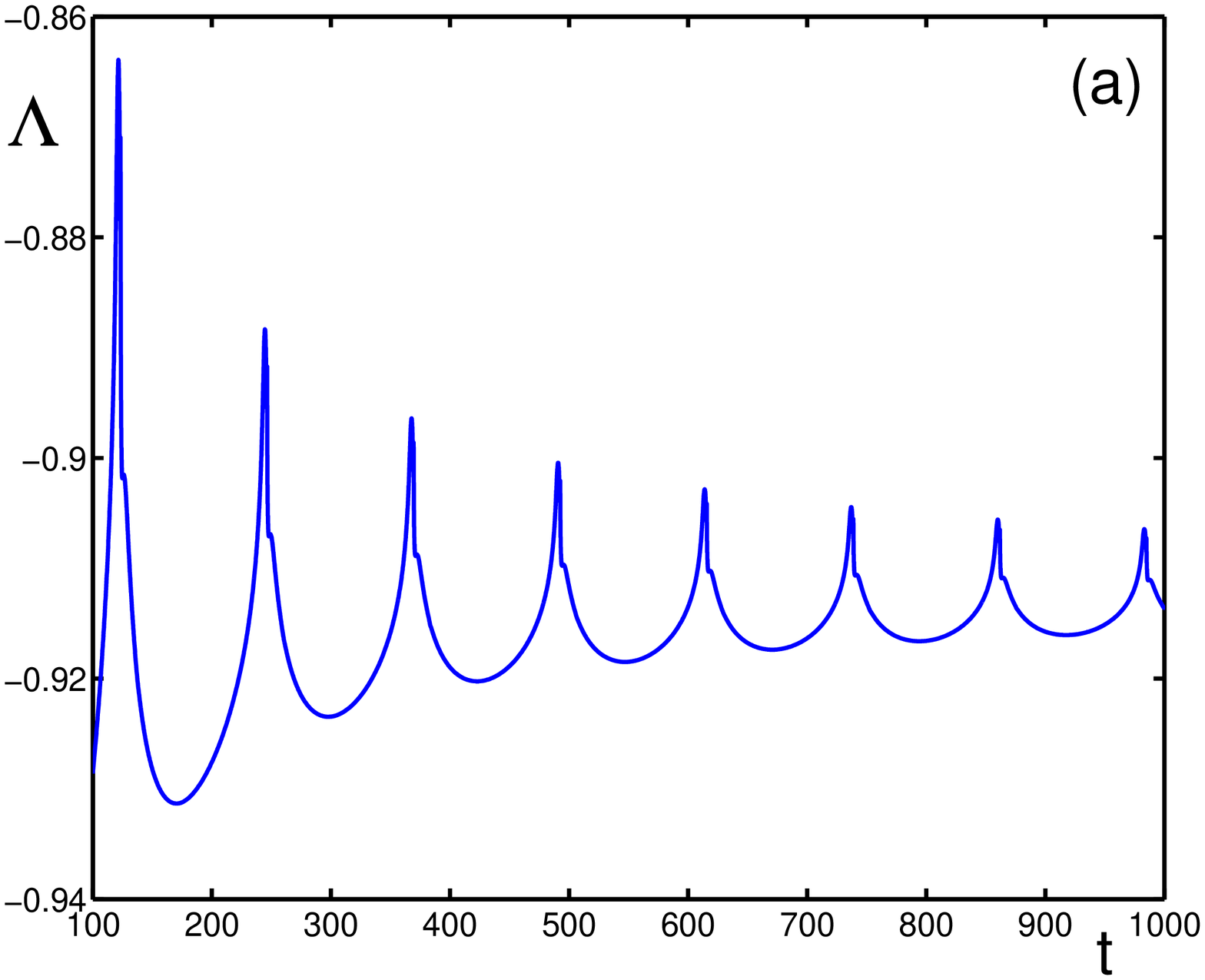} \hspace{1cm}
\includegraphics[width=8.5cm]{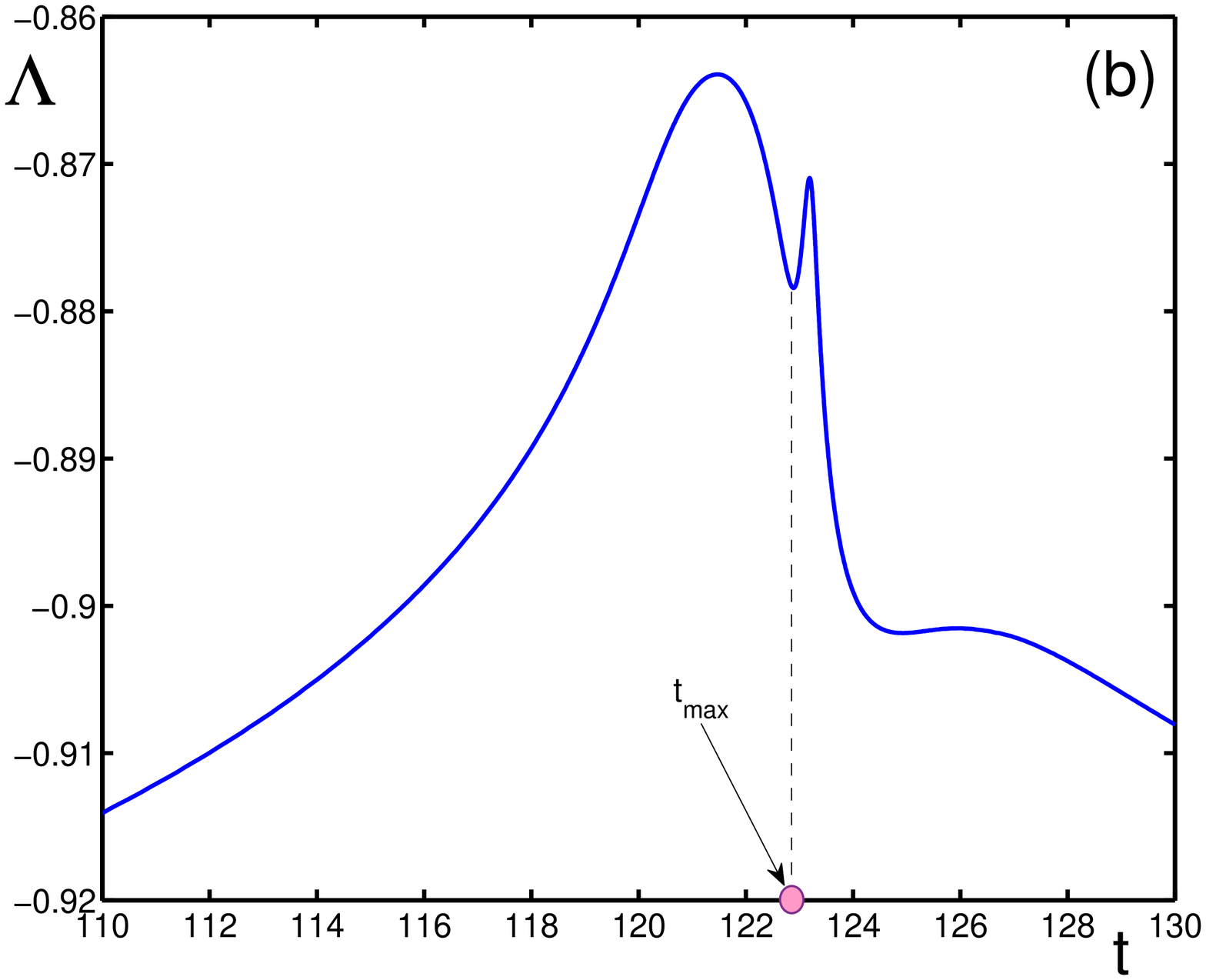} } }
\caption{Expansion exponent $\Lbd(t)$, with the parameters $b = 0.4$,
$g = -0.029$, and the initial conditions $x_0 = 1$ and $z_0 = 0.1$, for different
temporal scales: (a) $t \in [100,1000]$; (b) $t \in [110,130]$. At
$t_{max}\approx 122.89$, the asset price $x(t)$ has its first local maximum
$x_{max} = 61.7171$, which corresponds to the first local minimum of $\Lbd(t)$.
}
\label{fig:Fig.12}
\end{figure}

\newpage

\begin{figure}[ht]
\vspace{9pt}
\centerline{
\hbox{ \includegraphics[width=8.5cm]{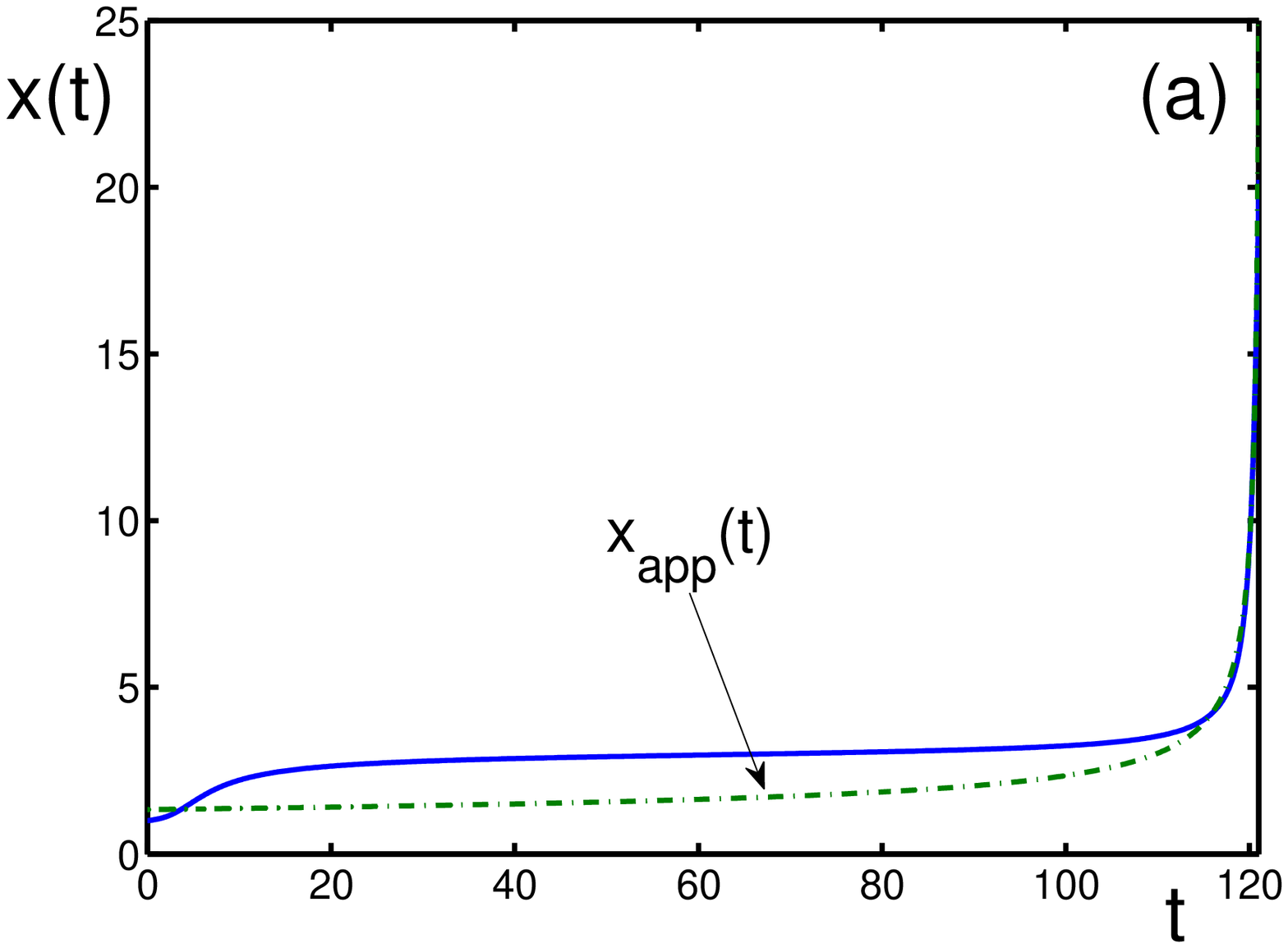} \hspace{1cm}
\includegraphics[width=8.5cm]{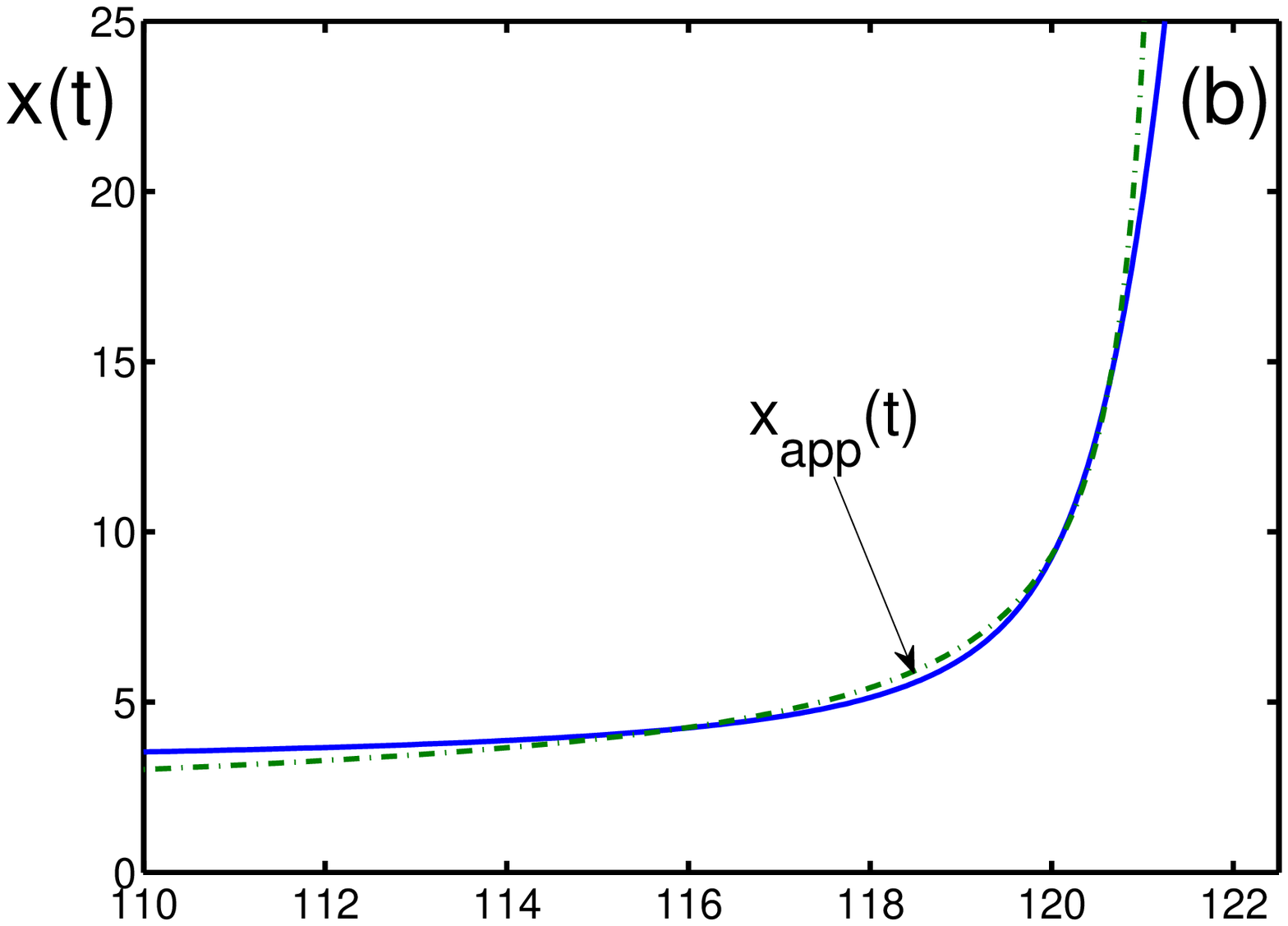}  } }
\vspace{9pt}
\centerline{
\hbox{ \includegraphics[width=7.5cm,height=5.5cm]{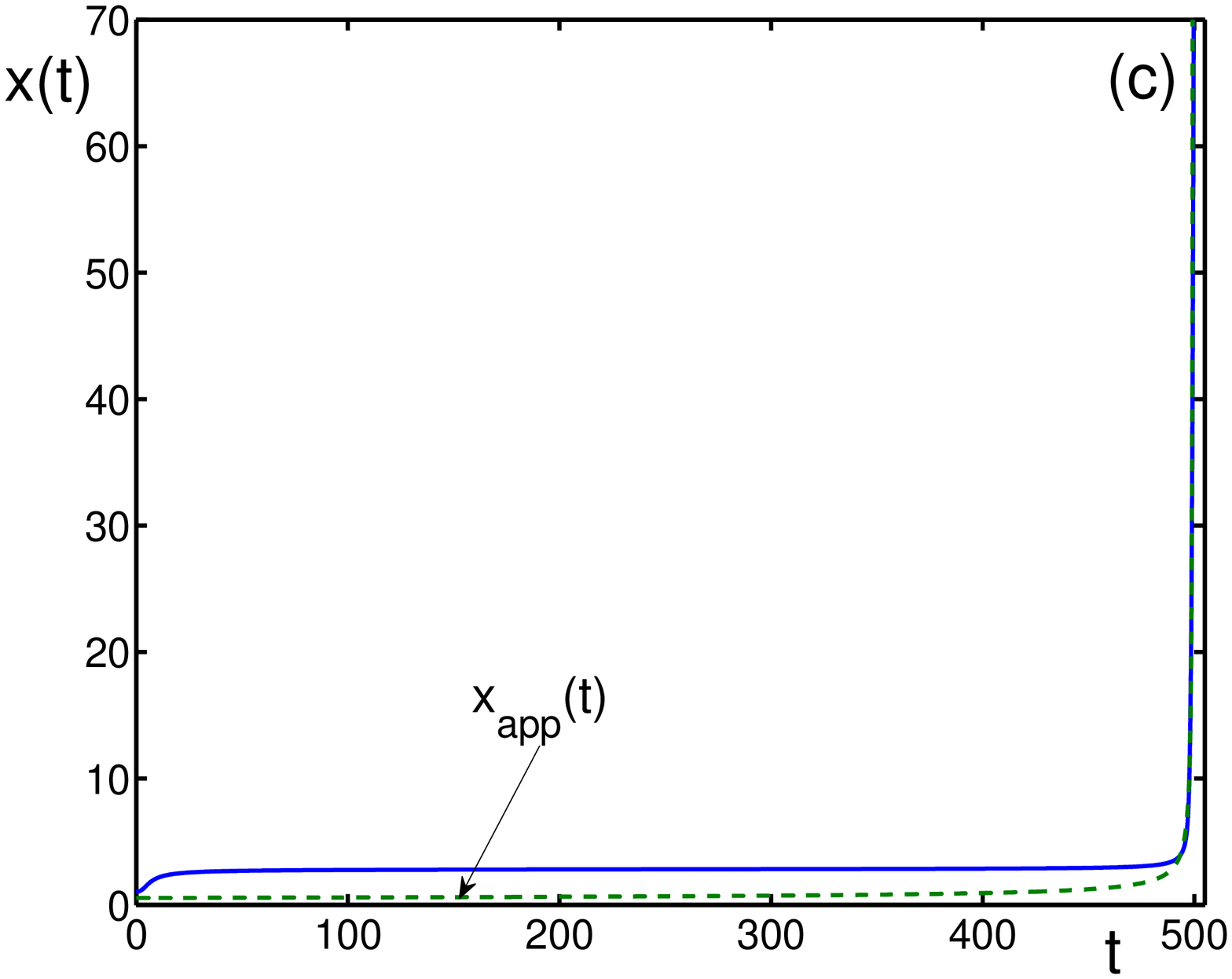}
\hspace{1.8cm}
\includegraphics[width=7.5cm,height=5.5cm]{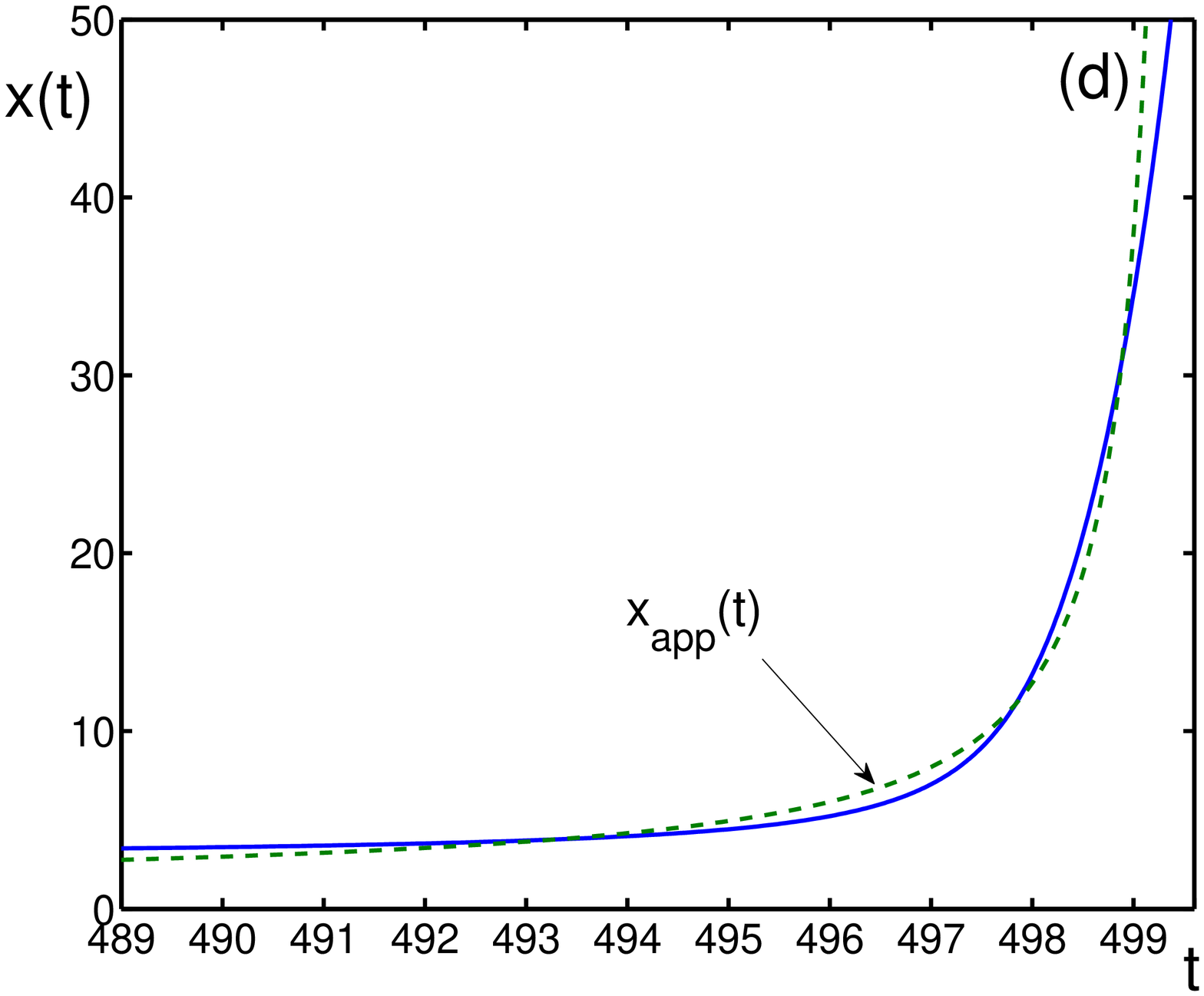} } }
\caption{Comparison of the numerical solution for the asset price
$x(t)$ (solid line) and its approximation $x_{app}(t)$ (dashed-dotted line),
with initial conditions $x_0 = 1$ and $z_0 = 0.1$, for different parameters
$b$ and $g$, where $1/e < b < b_0 \approx 0.47$, and for different intervals 
of time: (a) $b = 0.4$, $g = -0.029 > g_c = -0.02943$, and $t \in [0,121]$;
(b) $b$ and $g$, as in Fig. 13a, but $t \in [110,121]$; (c) $b = 0.38$, 
$g = -0.0117 > g_c = -0.01174$, and $t \in [0,499.6]$; (d) $b$ and $g$, as
in Fig. 13c, but $t \in [489,499.6]$;
}
\label{fig:Fig.13}
\end{figure}

\newpage

\begin{figure}[ht]
\vspace{9pt}
\centerline{
\hbox{ \includegraphics[width=8.5cm]{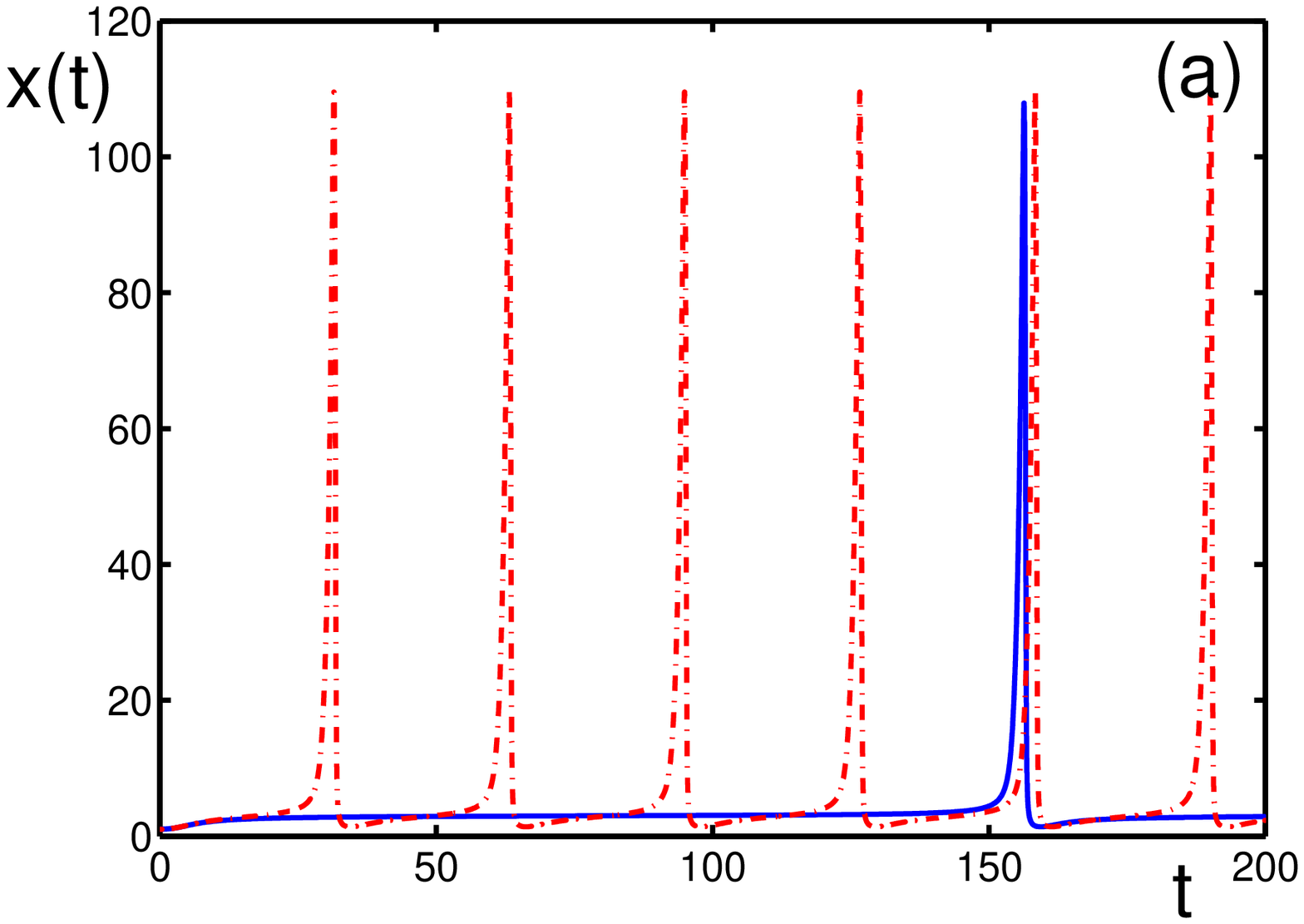} \hspace{1cm}
\includegraphics[width=8.5cm,height=6cm]{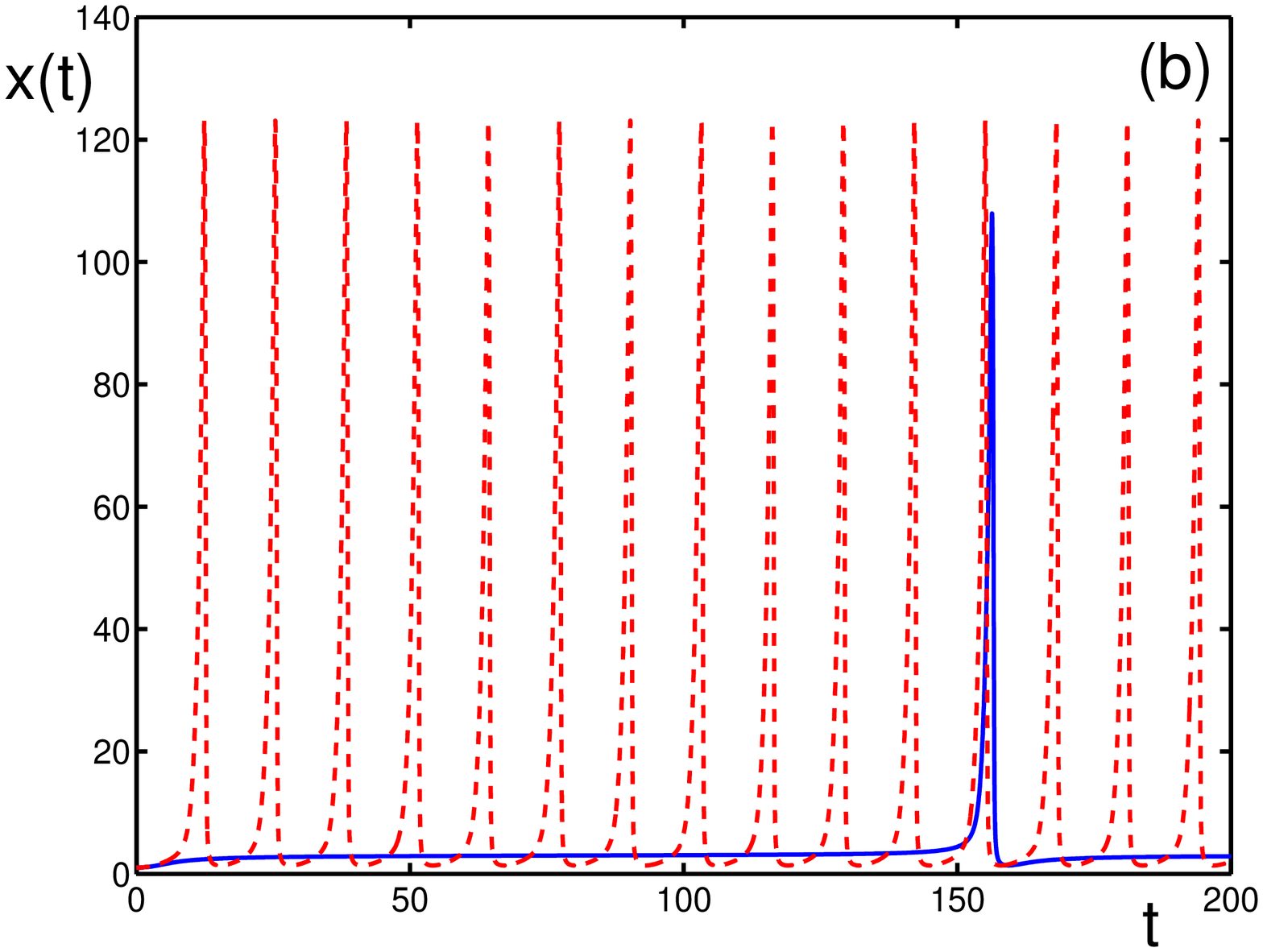}  } }
\vspace{9pt}
\centerline{
\hbox{ \includegraphics[width=8.5cm]{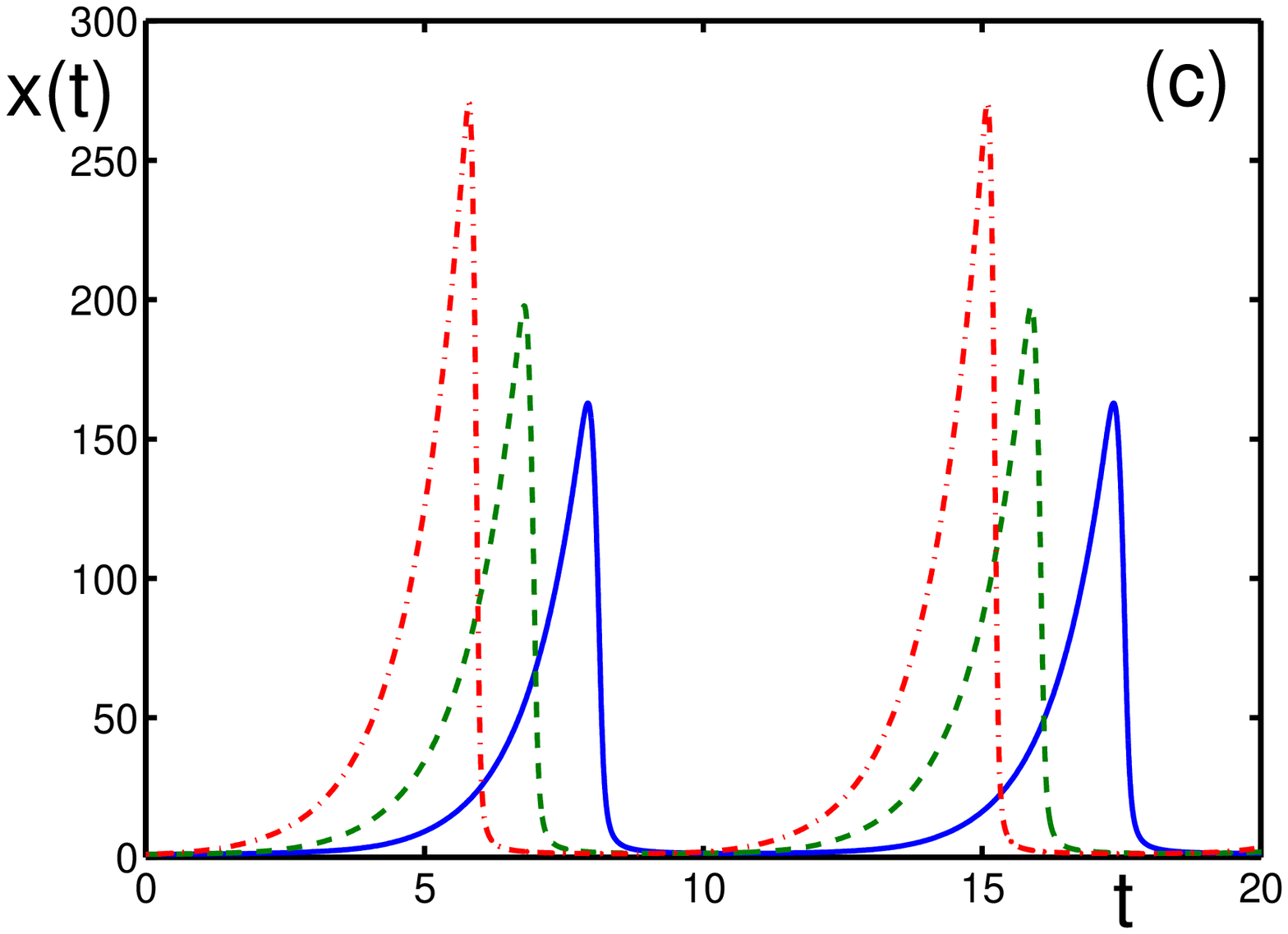} } }
\caption{Change in the behavior of the asset price $x(t)$, with fixed
$g = -0.03$, when varying $b$ in the vicinity of the boundary point
$b_2 = 0.400691$ separating the regions with different numbers of fixed
points. The initial conditions are $x_0 = 1$ and $z_0 = 0.1$.
(a) Asset price $x(t)$ for $b = 0.401 > b_2$ (solid line) and for $b = 0.41$
(dashed-dotted line) for $t \in [0,200]$; (b) asset price $x(t)$ for
$b = 0.401 > b_2$ (solid line) and for $b = 0.501$ (dashed-dotted line) for
$t \in [0,200]$; (c) asset price $x(t)$ for $b = 1$ (solid line), for $b = 2$
(dashed line), and for $b = 10$ (dashed-dotted line), with $t \in [0,20]$.
}
\label{fig:Fig.14}
\end{figure}

\newpage

\begin{figure}[ht]
\vspace{9pt}
\centerline{
\hbox{ \includegraphics[width=8.5cm]{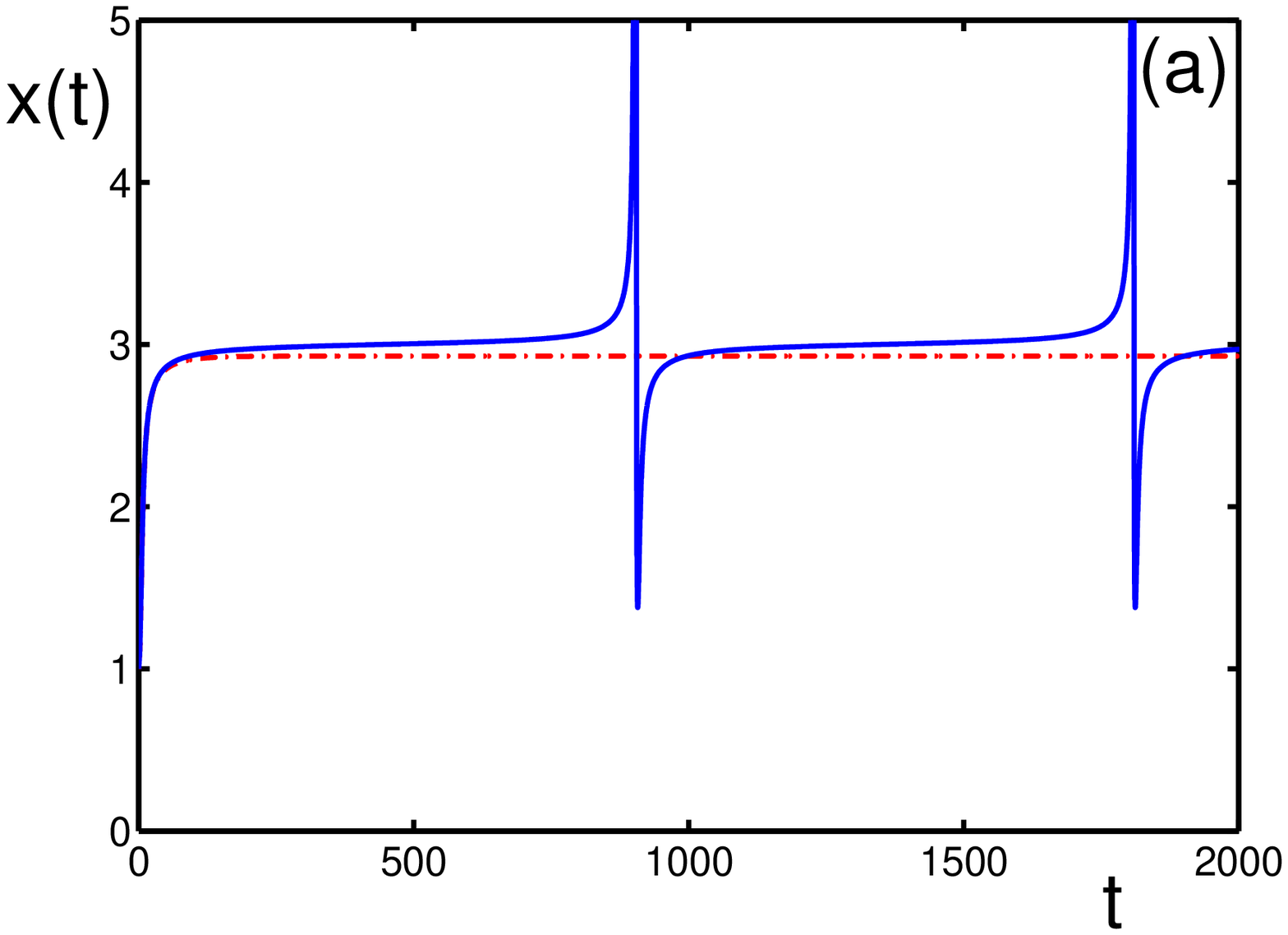} \hspace{1cm}
\includegraphics[width=8.5cm]{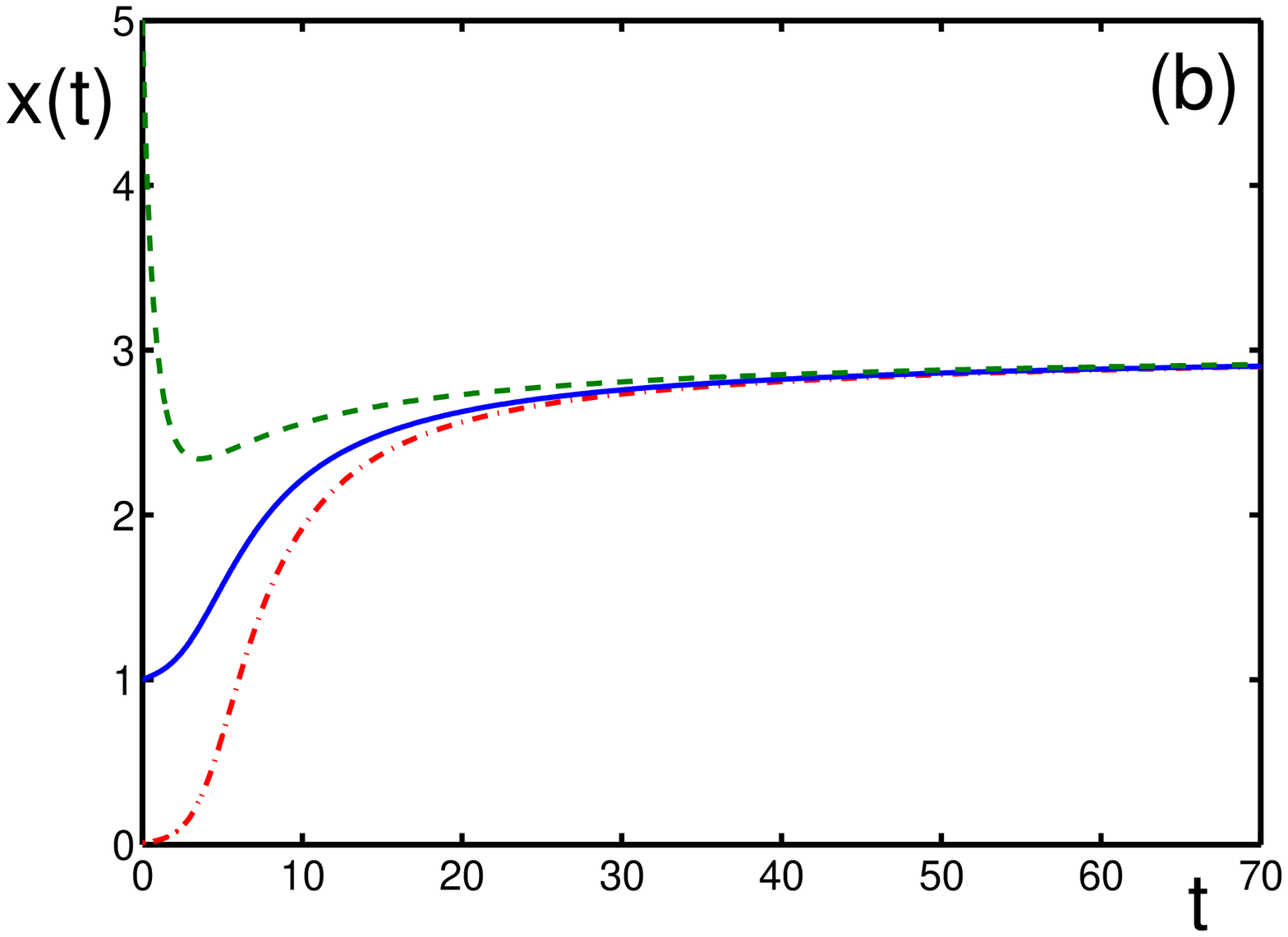} } }
\caption{Asset price $x(t)$ for the parameter $g = -0.03$, but with
different $b$ and for different initial conditions $\{x_0,z_0\}$.
(a) Asset price $x(t)$ for $b = 0.4007 > b_2$ (solid line), $b = 0.4006 < b_2$
(dashed-dotted line), and the initial conditions $x_0 = 1$ and $z_0 = 0.1$.
Here, $b_2 = 0.400691$ is the boundary point, such that for $b > b_2$, there
is an unstable focus and a limit cycle, while for $b < b_2$, there are three
fixed points. The fixed point $\{x_3 = 2.928, z_1 = 0.916\}$ is a stable node,
with the Lyapunov exponents $\{\lbd_1 = -0.903,\lbd_2 = -0.022\}$.
The fixed point $\{x_2 = 3.08, z_2 = 0.912\}$ is a saddle, with the Lyapunov
exponents $\{\lbd_1 = -0.899,\lbd_2 = +0.022\}$, and the fixed point
$\{x_1 = 58.26, z_1 = 0.174\}$ is an unstable focus, with the Lyapunov
exponents $\{\lbd_1 = \lbd_2^* = +1.03 - i1.72\}$. (b) Asset price $x(t)$
for the parameters $g = -0.03$, $b = 0.4007 > b_2$, and different initial
conditions: $\{x_0 = 1, z_0 = 0.1\}$ (solid line), $\{x_0 = 5, z_0 = 0.5\}$
(dashed line), and $\{x_0 = 0.01, z_0 = 50\}$ (dashed-dotted line).
}
\label{fig:Fig.15}
\end{figure}

\newpage

\begin{figure}[ht]
\vspace{9pt}
\centerline{
\hbox{ \includegraphics[width=8.5cm]{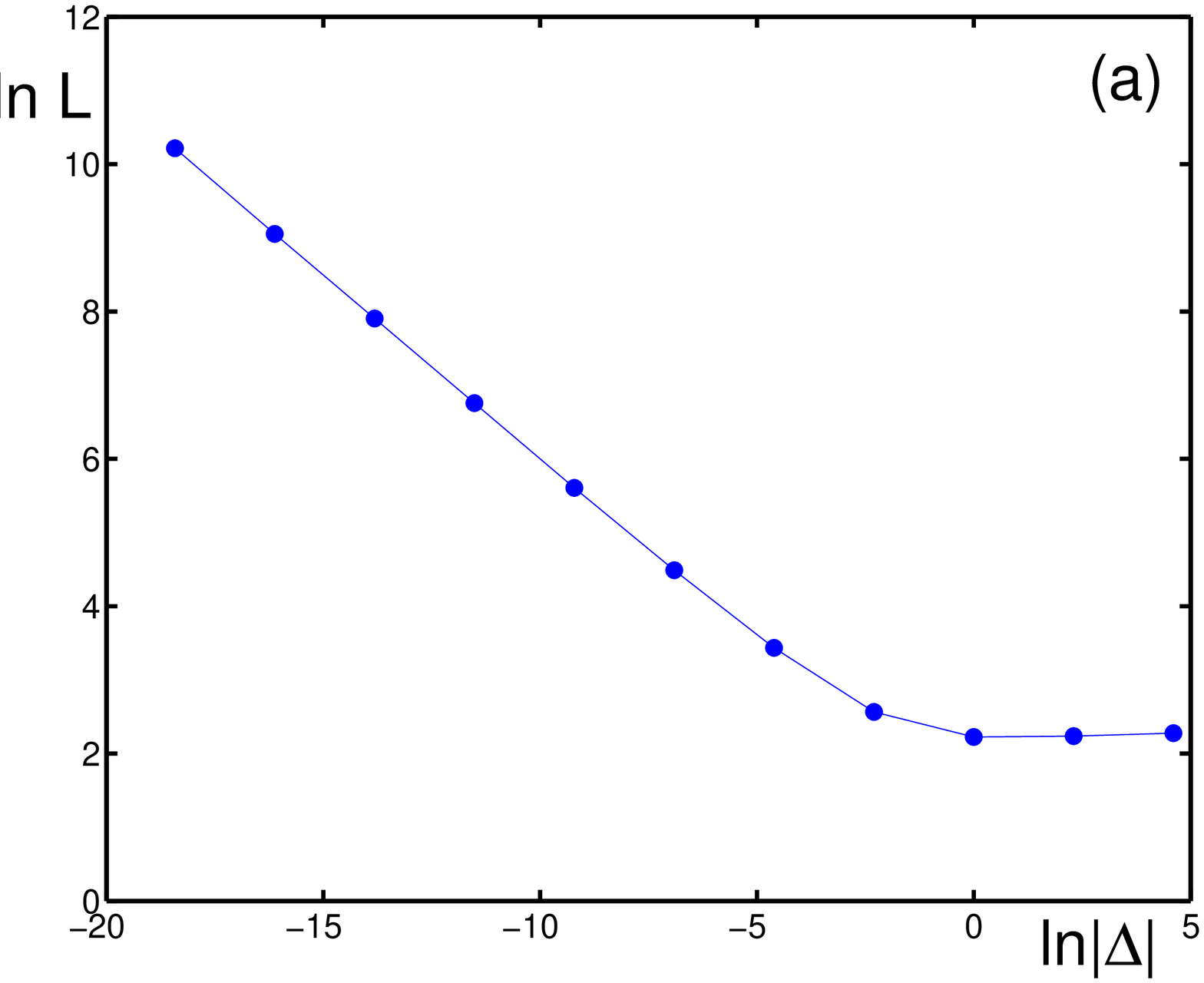} \hspace{1cm}
\includegraphics[width=8.5cm]{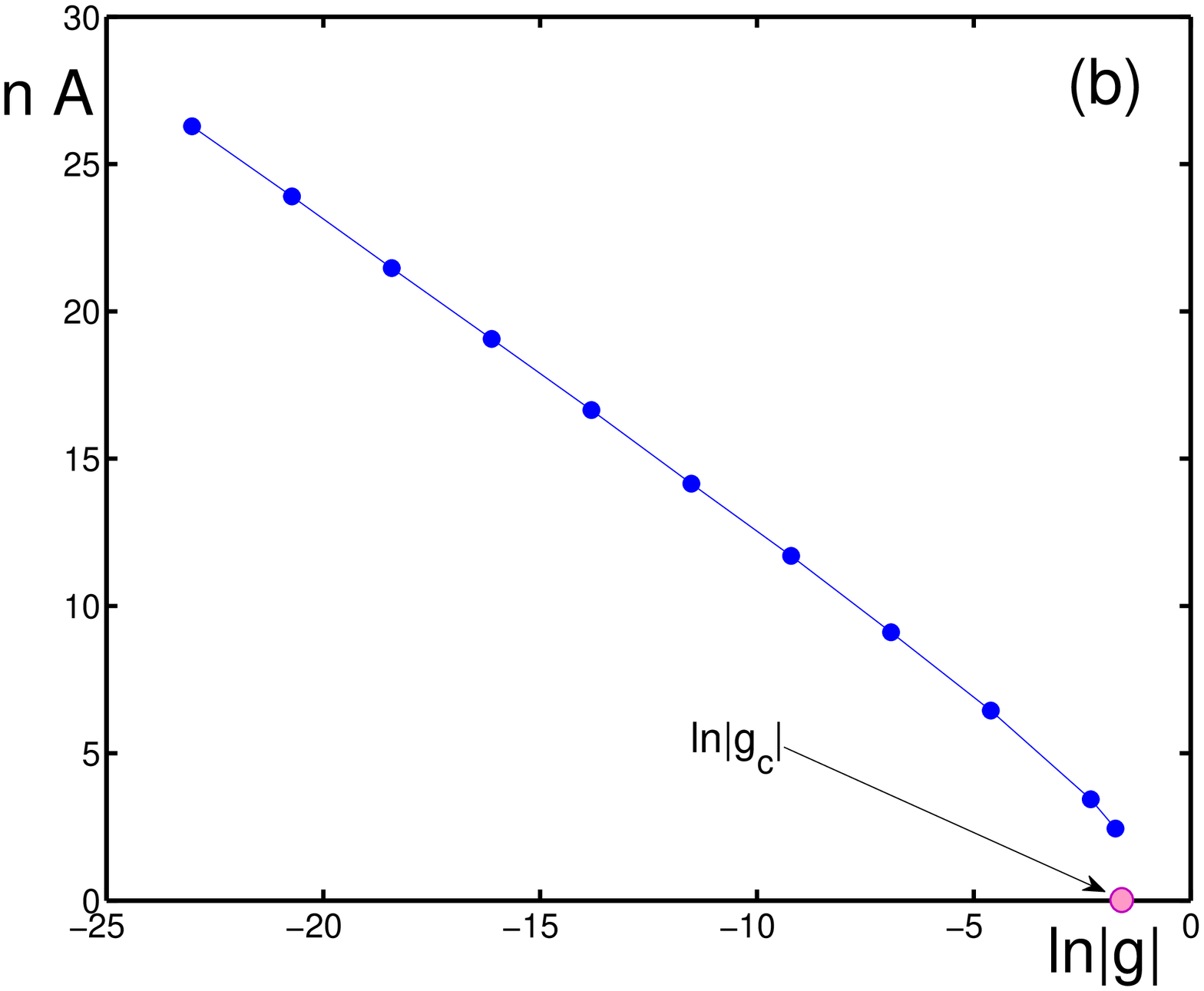} } }
\caption{Critical behavior of bubble characteristics. The temporal interval
between two neighboring bubbles is $L$. The bubble amplitude is $A$. The
distance from the critical line is $\Dlt = b - b_2$ in (a) and $g$ in (b).
The initial conditions are taken as $x_0 = 1$ and $z_0 = 0.1$.
(a) Interval $L$ in logarithmic units, as a function of log-distance
$\ln |\Dlt| = \ln |b - b_2|$ to the critical line. Here, $g = - 0.03$ and
$b_2 = 0.400691$. (b) Bubble amplitude $A$ in logarithmic units, as a
function of $\ln|g|$ measuring the distance from the critical line $g = -0$,
with fixed $b = 1$. Here $g_c = -0.1769$, so that $\ln g_c = -1.7322$.
}
\label{fig:Fig.16}
\end{figure}

\end{document}